\documentclass[twocolumn,10pt,journal,letterpaper]{IEEEtran}
\IEEEoverridecommandlockouts
\usepackage{algorithm}
\usepackage{graphicx}
\usepackage[noend]{algpseudocode}
\usepackage{url} 
\usepackage{etex} 
\usepackage{tikz}
\usetikzlibrary{shapes,arrows}
\usepackage{pstricks}
\usepackage{pst-node,pst-blur}
\usetikzlibrary{arrows}
\usepackage{amsfonts}
\usepackage{tabularx}
\usepackage{subfigure}
\usepackage{amssymb}
\usepackage{steinmetz}   
\usepackage{booktabs} 
\usepackage{multirow} 
\usepackage{epsfig}
\usepackage{epstopdf}
\usepackage{array}

\usepackage[noadjust]{cite}

\newtheorem{remark}{Remark}

\newtheorem{proposition}{Proposition}
\newtheorem{lemma}{Lemma}

\usepackage[cmex10]{amsmath}
\usepackage[cmex10]{mathtools}

\newcommand\abs[1]{\left\lvert#1\right\rvert}
\newcommand\norm[1]{\left\lVert#1\right\rVert}

\algdef{SE}[DOWHILE]{Do}{doWhile}{\algorithmicdo}[1]{\algorithmicwhile\ #1}%

\allowdisplaybreaks   

\makeatletter 
\def\@eqnnum{{\normalsize \normalcolor (\theequation)}} 
\makeatother

\begin{document} 
\title{Sum Throughput Maximization in Multi-Tag Backscattering to Multiantenna Reader}
\author{Deepak Mishra,~\IEEEmembership{Member,~IEEE,} and Erik G. Larsson,~\IEEEmembership{Fellow,~IEEE}
\thanks{D. Mishra and E. G. Larsson are with the Communication Systems Division of the Department of Electrical Engineering (ISY) at the Link\"oping University, 581 83 Link\"oping, Sweden (emails: \{deepak.mishra, erik.g.larsson\}@liu.se).}
\thanks{This work is supported in part by the Swedish Research Council (VR) and ELLIIT.}
\thanks{A preliminary  version~\cite{ICASSP19-Multitag} of this work was presented at the IEEE ICASSP, Brighton, UK, May 2019.}}
\maketitle

\begin{abstract} 	
Backscatter communication (BSC) is being realized as the core technology for pervasive sustainable Internet-of-Things applications. However, owing to the resource-limitations of passive tags, the efficient usage of multiple antennas at the reader is essential for both downlink excitation and uplink detection. This work targets at maximizing the achievable sum-backscattered-throughput by jointly optimizing the transceiver (TRX) design at the reader and backscattering coefficients (BC) at the tags. Since, this joint problem is nonconvex, we first present individually-optimal designs for the TRX and BC. We show that with precoder and {combiner} designs at the reader respectively targeting downlink energy beamforming and uplink Wiener filtering operations, the BC optimization at tags can be reduced to a binary power control problem. Next, the asymptotically-optimal joint-TRX-BC designs are proposed for both low and high signal-to-noise-ratio regimes. Based on these developments, an iterative low-complexity algorithm is proposed to yield an efficient jointly-suboptimal design. Thereafter, we discuss the practical utility of the proposed designs to other  application settings like wireless powered communication networks and BSC with imperfect channel state information. {Lastly, selected numerical results, validating the analysis and shedding novel insights, demonstrate that the proposed designs can yield significant  enhancement  in the sum-backscattered throughput over  existing benchmarks.}       
\end{abstract} 
 

\section{Introduction and Background}\label{sec:intro}
Backscatter communication (BSC) technology, comprising low-cost tags, without any bulkier radio frequency (RF) chain components, has gained significant recent attention owing to its potential in realizing the sustainable and pervasive ultra-low-power networking~\cite{New-SP-Mag2}. The key merit of BSC, not requiring any signal modulation, amplification, or retransmission by the tags, is that it shifts the high cost and large form-factor constraints to the reader side, leading to the tag-size miniaturization, which is the basic need of numerous smart networking applications~\cite{LiveTag}. Despite these potential merits, the widespread utility of BSC is limited by shorter read-range~\cite{Bistatic-BCS} and lower achievable data rates~\cite{BSC-Cascaded}. {Further, since the tags are  lightweight, passive, chipless, and battery-free devices that do not have their own radio circuitry to process incoming signals or estimate the channel response, multiple antennas at the reader are required to separate out the backscattered signals from  multiple tags by exploiting spatial  multiplexing to enhance data rate and BSC reliability. Also, this multiantenna reader can implement energy beamforming (EB) during the carrier transmission to significantly improve BSC range.} Therefore,  to enable efficient BSC from multiple tags, there is a need  for investigating the novel jointly-optimal transmit (TX) and receive (RX) beamforming at the multiantenna reader and backscattering designs at the  tags.
 
\subsection{State-of-the-Art}\label{sec:rw} 
BSC is based on the decoding of backscattered information signals at the reader as received from the multiple low-power tags. These tags  communicate their information to the reader by respectively modulating their load impedances to control the strength, phase, frequency, or any other characteristics of the carrier signal(s) as received and reflected back to the reader. {Depending on the energy constraints of the tags, BSC models can be divided into three groups: (a) \textit{passive}~\cite{QAM-BSC}, (b) \textit{semi-passive}~\cite{semi-passive}, and (c) \textit{active}~\cite{Mag-Amb-BSC}.}  Though both passive and semi-passive tags depend on the carrier signal excitation from the reader, the latter are also equipped with an internal power source to enable better reliability and longer range of accessibility. Whereas, the active tags are battery-powered and can broadcast their own signal, thereby achieving  much longer high link quality read range at the cost of bulkier size and higher maintenance requirements.    Similarly,  based on the network configuration, three main types of BSC models are:
\begin{itemize}
	\item \textit{Monostatic}: With carrier emitter and backscattered signal reader being the same entity, this model can share antennas for  transmission to and reception from tags~\cite{inv-BSC-MIMO}.   
	\item  \textit{Bi-static:}  Here, the emitter and reader are geographically-separated two different entities~\cite{Bistatic-BCS}. This
	model can help in achieving a longer range.
	\item  \textit{Ambient:} Widely investigated model where emitter is an uncontrollable source and the dedicated reader  decodes the resulting   backscattered information from the tags~\cite{Mag-Amb-BSC}. 
\end{itemize} 

\begin{figure}[!t]
\centering 
\includegraphics[width=3.4in]{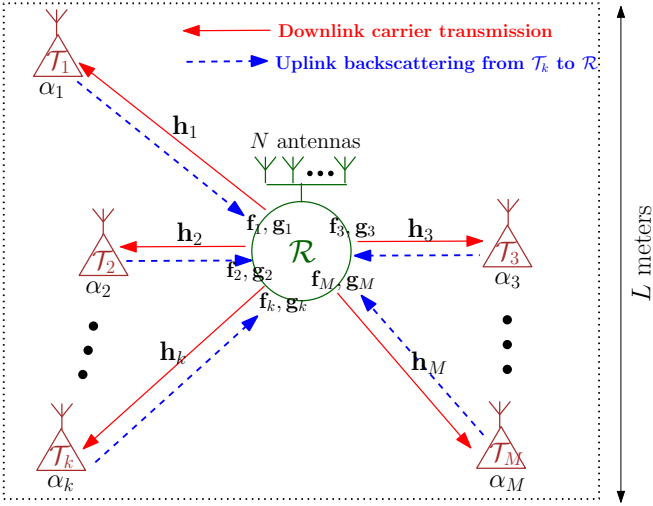}
\caption{\small Monostatic reciprocal-BSC channel model with full-duplex multiantenna $\mathcal{R}$ and multiple single-antenna tags.}\label{fig:model} 
\end{figure}

{{As a consequence, monostatic configurations are cheaper because they require relatively smaller number of antenna elements due to their sharing in full-duplex settings.} In contrast, the bi-static architectures ones  can  achieve   longer  read-range at the expense of combined higher antenna count for emission and reading purposes due to the geographic-separation of emitter and reader.} As shown in Fig.~\ref{fig:model}, we investigate a monostatic BSC system with multiple single-antenna semi-passive tags and a multiantenna reader working in  full-duplex mode~\cite{inv-BSC-MIMO,SPAWC18}. Henceforth, each antenna element at the reader is used for both carrier emission and backscattered signal reception~\cite{FD-Just}. {This adopted configuration with a large antenna array at reader can maximize the BSC range, while meeting the desired rate requirements of tags, by exploiting the array gains during the downlink carrier transmission to the multiple tags, and uplink multiplexing gains during backscattered signals reception at the reader. {Also, this setting is one of the most practical ones because it moves the computational-complexity and form-factor constraints from the low-power tags to a relatively-powerful reader.} Recent, experimental results~\cite{MIMO-RFID-Exp1,BackFi} have corroborated this fact that coverage range can be significantly improved up to a few hundred meters by exploiting array gains at reader.} However, these gains in the multiple-input-multiple-output (MIMO) reader-assisted BSC  can be strongly enhanced by optimally designing the underlying transceiver (TRX).

Noting that the tags-to-reader backscatter uplink channel is  coupled  to  the reader-to-tags downlink one, novel higher order modulation  schemes have been investigated in \cite{inv-BSC-MIMO} for the monostatic MIMO-BSC settings. Whereas, a frequency-modulated continuous-wave BSC system with monostatic reader, whose one antenna was dedicated for transmission and remaining for  the reception of backscattered signals, was studied in~\cite{RFID-TSP} to precisely determine the number and position of active tags. On similar lines, considering a multiantenna power beacon assisted bi-static BSC model, robust inference algorithms, not requiring any channel state or statistics information, were proposed in~\cite{zhu2017inference} to detect the sensing values of multiple single antenna backscatter sensors at a multiantenna reader by constructing Bayesian networks and using expectation maximization principle. Pairwise error probability and diversity order achieved by the orthogonal space time block codes over the dyadic backscatter channel (i.e., monostatic BSC system with multiple-antennas at the reader for transmission and reception from a multiantenna tag) were derived in \cite{inv-BSC-MIMO} and \cite{STC_BSC}. {Authors in~\cite{Amb-BCS-Ana} designed a data detection algorithm for
	an  ambient BSC system where differential  encoding was adopted at the tag to  eliminate  the  necessity  of channel estimation (CE) in minimizing the underlying sum bit error rate (BER) performance.  The asymptotic outage performance of  an adaptive ambient BSC scheme with Maximum  Ratio  Combining  (MRC) at the multiantenna reader was analyzed in~\cite{AmbBSC-MRC} to demonstrate its superiority over the traditional non-adaptive scheme. Adopting the BSC model with multiple antennas the reader, authors in \cite{GIoT-BSC} first presented maximum likelihood (ML) based optimal {combiner} for simultaneously recovering the signals from emitter and tag. Then, they also investigated the relative performance of the suboptimal linear {combiners} (MRC, Zero Forcing (ZF)  and  Minimum Mean-Squared Error (MMSE)) and successive interference cancellation (SIC) based {combiners}, where MMSE-SIC  {combiner} was shown to achieve the near-ML detection performance.} {In~\cite{MultiAntR1}, a dyadic backscatter channel between multiantenna tag and reader was studied to quantify the impact of underlying pin-hole  diversity and the  RF  tag's  scattering  aperture on enhancing the achievable BER performance and tag operating range. Authors in~\cite{MultiAntR2} noted that if separate 	reader transmitter and receiver antennas are used in conjunction with multiple RF tag antennas,  the envelope 	  correlation  between  the  forward  and  backscatter  links  can be significantly reduced to enhance the BER performance. Furthermore, investigating the optimal detection threshold for ambient BSC in~\cite{ChinaConf}, it was found that an increasing array size can  yield larger gains 	in  BER  at  low  signal-to-noise (SNR),  with lower returns  in  high  SNR regime.}

On different lines, with the goal of optimizing harvested energy among tags,  sub-optimal EB designs for monostatic multiantenna reader were investigated in~\cite{BSC-WET-MIMO}. {More recently, a least-squares-estimator for the BSC channels between a multiantenna reader and single-antenna tag was proposed in~\cite{SPAWC18}. Based on that a linear MMSE based channel estimator was designed in~\cite{TSP19}  to come up with an optimal energy allocation scheme  maximizing underlying single-tag BSC performance while optimally selecting number of orthogonal pilots for CE.}
 

\subsection{Notations Used}    
The vectors and matrices are respectively denoted by boldface lower-case and capital letters. $\mathbf{A}^{\mathrm H}$, $\mathbf{A}^{\mathrm T}$, and $\mathbf{A}^{\mathrm *}$ respectively denote the Hermitian transpose, transpose, and conjugate of matrix $\mathbf{A}$. $\mathbf{0}_{n\times n},\mathbf{1}_{n\times n},$ and $\mathbf{I}_{n}$ respectively represent  $n\times n$ zero, all-ones, and identity matrices.  $[\mathbf{A}]_{i,j}$ stands for   $(i,j)$-th element of matrix $\mathbf{A}$, $[\mathbf{A}]_{i}$ represents for   $i$-th column of  $\mathbf{A}$, and $[\mathbf{a}]_{i}$ stands for  $i$-th element of vector $\mathbf{a}$. With $\mathrm{Tr}\left(\mathbf{A}\right)$ and  $\mathrm{rank}\left(\mathbf{A}\right)$ respectively being the trace and rank of matrix $\mathbf{A}$,  $\lVert\,\cdot\,\rVert$ and $\left|\,\cdot\,\right|$ respectively represent Frobenius norm of a complex matrix and absolute value of a complex scalar. ${\rm diag}\{\mathbf{a}\}$ is used to denote a square diagonal matrix with $\mathbf{a}$'s elements in its main diagonal and ${\rm vec}\{\mathbf{A}\}$ for representing the vectorization of matrix $\mathbf{A}$  into a column vector. $\mathbf{A}^{-1}$ and $\mathbf{A}^{1/2}$ represent the inverse and square-root, respectively, of a square matrix $\mathbf{A}$, whereas $\mathbf{A}\succeq0$ means that $\mathbf{A}$ is positive semidefinite and operator $\odot$ represents the
Hadamard product of two matrices. Expectation is defined using $\mathbb{E}\left\lbrace\cdot\right\rbrace$ and $\mathrm{v}_{\max}\left\lbrace\mathbf{A}\right\rbrace$ represents the principal eigenvector corresponding to maximum eigenvalue $\lambda_{\max}\left\lbrace\mathbf{A}\right\rbrace$ of {a Hermitian matrix}  $\mathbf{A}$. Lastly, with $j=\sqrt{-1}$, $\mathbb{R}$ and  $\mathbb{C}$ respectively denoting the  real and complex number sets, $\mathbb{C} \mathbb{N}\left(\boldsymbol{\mu},\mathbf{C}\right)$ denotes the complex Gaussian distribution with mean $\boldsymbol{\mu}$ and covariance  $\mathbf{C}$. 


\section{Motivation and Significance}\label{sec:motiv}
This section first discusses the novel aspects of this work targeted towards addressing an important existing research gap along with the potential scope of the proposed designs. In the latter half, we summarize the main contributions of this paper.  

\subsection{Novelty and Scope}\label{sec:novel} 

Since the information sources in BSC, i.e., tags, do not have their own RF chains for communication, the two key roles of the reader are: (a) carrier transmission to excite the tags in the downlink, and
(b) efficient detection of the received backscattered signals in the uplink. Therefore, new TRX designs are needed because the requirements of RX design for the uplink involving effective detection of the  backscattered information signals at the reader as received from the multiple tags are different from those of the TX beamforming in the downlink involving single-group multicasting-based carrier transmission. {Furthermore, the underlying nonconvex optimization problem is more challenging than in conventional wireless networks because the corresponding backscattered throughput definition involves product or cascaded channels. Also, the resource-limitations of tags put  additional constraints on the precoder and {combiner} designs.} 

{The existing works~\cite{FD-Just,RFID-TSP,MIMO-RFID-Exp1,BackFi,zhu2017inference,inv-BSC-MIMO,STC_BSC,GIoT-BSC,BSC-WET-MIMO,SPAWC18,TSP19,ChinaConf,Retro-BSC}  on multiantenna reader supported BSC did not focus on utilizing reader's efficacy in designing smart signal processing techniques  to overcome the radio limitations of  tags by jointly exploiting the array and multiplexing gains.} To the best of our knowledge, \textit{the joint TRX design for the multiantenna reader} has not been investigated in the literature yet. Also, the \textit{backscattering coefficient (BC) optimization at the tags} for maximizing the sum-backscattered-throughput is missing in the existing state-of-the art on the \textit{multi-tag  BSC systems}. Recently, a few BC design policies were investigated in~\cite{Retro-BSC} for maximizing the average harvested power due to the retro-directive beamforming at multiantenna energy transmitter based on the backscattered signals from the multiple single antenna tags. But~\cite{Retro-BSC} ignored the possibility of uplink backscattered information transfer, and only focused on the downlink energy transfer. 

{{In this work, we have presented {novel design insights} for both   TRX and BC optimization.}} Specifically, new solutions for  the  {individually-optimal} designs and  {asymptotically-global-optimal} joint-designs are proposed along with an efficient \textit{low-complexity} iterative algorithmic implementation. These designs can meet the basic requirement of extending the BSC range and coverage by imposing the non-trivial smart signal  processing at multiantenna reader. Significance of the proposed designs is corroborated  by the fact that they can yield substantial gains without relying on any assistance from the resource-constrained tags in solving the underlying nonconvex sum-backscattered-throughput maximization problem. {Our optimal designs are targeted for serving applications with the overall BSC system-centric goal, rather than individual tag-level, where the best-effort delivery is desired to maximize the aggregate throughput. Practical utility of these designs targeted for  monostatic BSC can be easily extended for addressing the needs of other BSC models. Also, we discourse how the proposed optimization techniques can be used for solving the  nonconvex throughput maximization problems in wireless powered communication networks (WPCN). Thus, this investigation, providing designs for achieving longer read-range  and higher backscattered-throughput,  enables widespread applicability of  BSC technology in  ultra-low-power emerging-radio networks for last-mile connectivity and Internet-of-Things networking.}

\subsection{Key Contributions and Paper Organization}\label{sec:contib}  
Five-fold contribution of this work is summarized below.
\begin{itemize}
\item {A novel optimization framework has been investigated for  maximizing the sum-backscattered-throughput from  multiple single-antenna tags in a monostatic BSC setting. It involves: (i) smart allocation of reader's resources by optimally designing the  TRX, and (ii) maximizing the benefit of tags cooperation by optimally designing their BC.} The corresponding basic building blocks and problem definition addressed are presented in Section~\ref{sec:model}.
\item {Noting non-convexity of this joint problem,  we propose new individually-optimal designs for TX precoding, RX beamforming at reader, and BC at the tags in Section~\ref{sec:SRM}. Further, their respective generalized-convexity~\cite{Baz} is explored along with individual global-optimality.
\item Next, the asymptotically-optimal joint designs are derived in Sections~\ref{sec:high} and~\ref{sec:low} for the high and low SNR regimes, respectively. We show that  both these jointly-optimal designs, which can be efficiently obtained, provide key novel design insights.  Using these results as performance bounds, a low-complexity iterative-algorithm is outlined   in Section~\ref{sec:Alg} to obtain a near-optimal  design.} 
\item To corroborate the practical utility of the proposed designs, in Section~\ref{sec:dis}  we discuss their extension to address the requirements of application networks like WPCN, bi-static and ambient BSC models with imperfect channel state information (CSI) and multiantenna tags. 
\item Detailed numerical investigation is carried out in Section~\ref{sec:res} to validate the analytical claims, present key optimal design insights, and quantify the performance gains over the conventional designs. {There other than comparing the efficacy of  individually-optimal designs, we have shown that on an average $>20\%$ sum-throughput gains can be achieved by the proposed joint TRX and BC design over the relevant benchmarks~\cite{WPCN-MIMO-SRM,GIoT-BSC}.}  
\end{itemize}
Throughout this paper, the main outcomes have been highlighted as remarks and Section~\ref{sec:concl} concludes this work with the keynotes and possible future research extensions.

\section{Problem Definition}\label{sec:model}
We start with briefly describing the adopted system model and network architecture, followed up by the BSC and semi-passive tag models. Later, we   present the expression for the achievable backscattered-throughput at reader from each tag.

\subsection{System Model and Network Architecture}\label{sec:sysMod}
We consider a multi-tag monostatic BSC system comprising $M$ single-antenna semi-passive tags, and one full-duplex reader equipped with $N$ antennas  which is responsible for simultaneous carrier transmission and backscattered signal decoding. Hereinafter, the $k$-th tag is denoted by $\mathcal{T}_k$ with $ k\in\mathcal{M}\triangleq \{1,2,\ldots,M\},$ and the reader is denoted by $\mathcal{R}$.
We assume that these $M$ tags are randomly deployed in a square field of length $L$ meters (m), with $\mathcal{R}$ being at its center  as shown in Fig.~\ref{fig:model}.    To enable  full-duplex operation~\cite{FD-Just}, each of the $N$ antennas at $\mathcal{R}$ can  transmit a carrier signal to the tags while concurrently receiving the backscattered signals. 

The multiantenna $\mathcal{R}$ adopts linear precoding and assigns each $\mathcal{T}_k$ a dedicated precoding vector  $\mathbf{f}_k\in\mathbb{C}^{N\times1}$. We denote by $\mathbf{s}_{\mathcal{R}}\sim\mathbb{C} \mathbb{N}\left(\mathbf{0}_{M\times 1},\mathbf{I}_M\right)$  the vector of $M$ independent  and
identically  distributed (i.i.d.) symbols as simultaneously transmitted by $\mathcal{R}$.  Hence, the complex baseband transmitted signal from $\mathcal{R}$ is given by $\mathbf{x}_{\mathcal{R}}\triangleq\sum_{k\in\mathcal{M}}\mathbf{f}_k\left[\mathbf{s}_{\mathcal{R}}\right]_k\in\mathbb{C}^{N\times 1}$, and we assume that there exists a total power budget $P_T$ to support this transmission.  The resulting $M$ modulated reflected data symbols as simultaneously backscattered from the $M$ tags are respectively spatially separated by $\mathcal{R}$ with the aid of $M$ linear  decoding vectors as denoted by  $\mathbf{g}_1,\mathbf{g}_2,\ldots,\mathbf{g}_M\in\mathbb{C}^{N\times 1}$.  Here combiner $\mathbf{g}_k$ is used for decoding $\mathcal{T}_k$'s  message. 
This restriction on TRX designs to be linear has not only been considered to address low-power-constraints of BSC, but also because for $N\gg M$, these designs are nearly-optimal~\cite{massive-MIMO}.

\subsection{Adopted BSC and Tag Models}\label{sec:tag}
In contrast to the practical challenges in implementing the full-duplex operation in conventional communication systems involving modulated information signals, the unmodulated carrier leakage in monostatic full-duplex BSC systems can be efficiently suppressed~\cite{FD-Just}. {Further, we consider semi-passive tags~\cite{semi-passive} that utilize the RF signals from $\mathcal{R}$ for backscattering their information and are also equipped with an internal power source or battery to support their low power on-board operations. Thus, they do not have to wait for having enough harvested energy, thereby reducing their overall  access delay~\cite{Mag-Amb-BSC}. However, note that this  battery  is only used for powering the tag's circuitry  to set the desired  modulation or BC and for regular operations like sensing. Also, these benefits of longer BSC range and higher rate due to an on-tag battery suffer from  few problems like extra weight, larger size, higher cost, and battery-life constraints.}

For implementing the backscattering operation, we consider that each $\mathcal{T}_k$ modulates the carrier received from $\mathcal{R}$ via a complex baseband signal denoted by $\mathrm{x}_{\mathcal{T}_k}\triangleq A_k-\zeta_{(k)}$~\cite{Bistatic-BCS}. Here, the load-independent constant $A_k$ is related to the  antenna structure of the $k$th tag  and the load-controlled reflection coefficient $\zeta_{(k)}\in\{\zeta_1,\zeta_2,\ldots,\zeta_V\}$ switches between the $V$ distinct values to implement the desired tag modulation~\cite{BSC-IoT}. 
{Without the loss of generality, to produce impedance values realizable with passive components, we assume that the effective signal $\left[\mathbf{s}\right]_k\triangleq\frac{\mathrm{x}_{\mathcal{T}_k}\sqrt{\alpha_k}}{\abs{\mathrm{x}_{\mathcal{T}_k}}}$ from each tag $\mathcal{T}_k$  satisfies  $\mathbb{E}\left\lbrace\left[\mathbf{s}\right]_k^*\left[\mathbf{s}\right]_k\right\rbrace=\alpha_k\in\left[0,1\right]$ because {the scaling factor corresponding to the magnitude  $\abs{\mathrm{x}_{\mathcal{T}_k}}$ of the $\mathcal{T}_k$'s complex baseband signal $\mathrm{x}_{\mathcal{T}_k}= A_k-\zeta_{(k)}$ can be included in its reflection coefficient or BC $\alpha_k$ definition~\cite{QAM-BSC}.}} The higher values of $\alpha_k$ reflect increasing amounts of the incident RF power back to $\mathcal{R}$ which thus result in higher backscattered signal strength and thereby maximizing the overall read-range of  $\mathcal{R}$. Whereas, the lower value of BC for a tag implies that its backscattering to $\mathcal{R}$ causes lesser interference for the other tags.

The $\mathcal{T}_k$-to-$\mathcal{R}$ wireless reciprocal-channel is denoted by an $N\times1$ vector $\mathbf{h}_k\sim\mathbb{C} \mathbb{N}\left(\mathbf{0}_{N\times 1},\beta_k\,\mathbf{I}_N\right)$. Here, parameter $\beta_k$ represents average channel power gain incorporating the fading gain and propagation loss over $\mathcal{T}_k$-to-$\mathcal{R}$ or $\mathcal{R}$-to-$\mathcal{T}_k$ link. {Although we have considered i.i.d. fading coefficients $\mathbf{h}_k$ for all $\mathcal{T}_k$-to-$\mathcal{R}$ channels due to sufficient antenna separation at reader~\cite{STC_BSC,inv-BSC-MIMO,TSP19}, the proposed designs in this work can also be used for the BSC settings with dependent and not necessarily identically distributed fading scenarios. In this paper we assume that this perfect CSI for each $\mathbf{h}_k$ is available at $\mathcal{R}$ to investigate the best achievable performance. However, our proposed designs can be extended to imperfect CSI cases as discussed in Section~\ref{sec:imperfect-CSI} and their robustness under inaccuracy in CSI is also demonstrated in Section~\ref{sec:res}.}

Therefore, on using these models,  the baseband received signal $\left[\mathbf{y}_{\mathcal{T}}\right]_k$ at  $\mathcal{T}_k$ is expressed as:
\begin{equation}\label{Eq:System}
\left[\mathbf{y}_{\mathcal{T}}\right]_k=\mathbf{h}_{k}^{\rm T}\sum_{m\in\mathcal{M}}\mathbf{f}_m\left[\mathbf{s}_{\mathcal{R}}\right]_m + \left[\mathbf{w}_{\mathcal{T}}\right]_k,\quad\forall k\in\mathcal{M},
\end{equation}
where $\left[\mathbf{s}_{\mathcal{R}}\right]_k\sim\mathbb{C} \mathbb{N}\left(0,1\right)$ for each $\mathcal{T}_k$ are  i.i.d. symbols and $\mathbf{w}_{\mathcal{T}}$ is the zero-mean Additive White Gaussian Noise (AWGN) vector with independent entries having variance $\sigma_{\mathrm{w}_\mathcal{T}}^2$. 

\subsection{Backscattered-Throughput at $\mathcal{R}$}\label{sec:BSC-T}
{We note that the backscattered noise strength due to the AWGN power ${\sigma_{\mathrm{w}_\mathcal{T}}^2}$ is practically negligible~\cite{QAM-BSC,semi-passive,Mag-Amb-BSC,inv-BSC-MIMO,Bistatic-BCS,SPAWC18} in comparison to the corresponding carrier reflection strength due to the signal power $\sum_{m\in\mathcal{M}}\left|\mathbf{h}_{k}^{\rm T}\mathbf{f}_m\right|^2$. So, ignoring this backscattered noise, which in comparison to the excitation power gets practically lost during  backscattering from  tags, the received signal $\mathbf{y}_{\mathcal{R}}\in\mathbb{C}^{N\times 1}$ available for information decoding at $\mathcal{R}$,  as obtained using the definition \eqref{Eq:System}, is:
\begin{align}\label{eq:yi}
\mathbf{y}_{\mathcal{R}} \triangleq&\, \sum_{m\in\mathcal{M}}\mathbf{h}_{m}\, \left[\mathbf{s}\right]_m \left[\mathbf{y}_{\mathcal{T}}\right]_m + \mathbf{w}_{\mathcal{R}}\nonumber\\
\approx&\,\sum_{m\in\mathcal{M}} \,\mathbf{h}_{m}\,\left[\mathbf{s}\right]_m\,\mathbf{h}_{m}^{\rm T}\sum_{k\in\mathcal{M}}\mathbf{f}_k\left[\mathbf{s}_{\mathcal{R}}\right]_k + \mathbf{w}_{\mathcal{R}}, 
\end{align}
where the $N\times 1$ vector $\mathbf{w}_{\mathcal{R}}\sim\mathbb{C} \mathbb{N}\left(\mathbf{0}_{N\times 1},\sigma_{\mathrm{w}_\mathcal{R}}^2\,\mathbf{I}_N\right)$ represents the received zero-mean AWGN at $\mathcal{R}$ and $\sigma_{\mathrm{w}_\mathcal{R}}^2$ is the noise power spectral density.} Applying the linear detection at $\mathcal{R}$, the received signal $\mathbf{y}_{\mathcal{R}}$ can be separated into $M$ streams by multiplying it with detection matrix ${\mathbf{G}}\triangleq\left[{\mathbf{g}}_1\;\;{\mathbf{g}}_2\;\;{\mathbf{g}}_3\;\ldots\;{\mathbf{g}}_M\right]$ and the corresponding decoded information signal is:
\begin{align}\label{eq:yd1}
\widehat{\mathbf{y}}_{\mathcal{R}} \triangleq\mathbf{G}^{\rm H}\,\mathbf{y}_{\mathcal{R}}\in\mathbb{C}^{M\times 1}.
\end{align}  
As each of the $M$ streams can be decoded independently, the complexity of the above linear receiver is on the order of $M\abs{\mathcal{S}}$, where $\abs{\mathcal{S}}$ denotes the cardinality of the finite alphabet set of $\left[\mathbf{s}\right]_k$ for each $\mathcal{T}_k$. Thus, with $\mathcal{M}_k\triangleq\mathcal{M}\setminus{\{k\}}=\{1,2,\ldots,k-1,k+1,k+2,\ldots,M\}$, the $k$th element of $\widehat{\mathbf{y}}_{\mathcal{R}}$, to be used for decoding the backscattered message of $\mathcal{T}_k$, is:
\begin{align}\label{eq:yd2}
\left[\widehat{\mathbf{y}}_{\mathcal{R}}\right]_k =&\,\mathbf{g}_{k}^{\rm H}\,\mathbf{h}_{k}\,\mathbf{h}_{k}^{\rm T}\left[\mathbf{s}\right]_k\sum_{m\in\mathcal{M}}\mathbf{f}_m  \left[\mathbf{s}_{\mathcal{R}}\right]_m +\sum_{i\in\mathcal{M}_k}\mathbf{g}_{k}^{\rm H}\,\mathbf{h}_{i}\,\mathbf{h}_{i}^{\rm T}\left[\mathbf{s}\right]_i\nonumber\\
&\,\times\sum_{m\in\mathcal{M}}\mathbf{f}_m\, \left[\mathbf{s}_{\mathcal{R}}\right]_m + \mathbf{g}_{k}^{\rm H}\,\mathbf{w}_{\mathcal{R}},\,\forall k\in\mathcal{M}.
\end{align}   

Therefore, using \eqref{eq:yd2} and the backscattered message $\left[\mathbf{s}\right]_k$ definition from Section~\ref {sec:tag},  the resulting signal-to-interference-plus-noise-ratio (SINR) $\gamma_{\mathcal{R}_k}$ at $\mathcal{R}$ from each $\mathcal{T}_k$ is given by:
\begin{align}\label{eq:SINR}
\gamma_{\mathcal{R}_k} \triangleq\frac{\alpha_k\,\gamma_{\mathcal{T}_k} \left|\mathbf{g}_{k}^{\rm H}\,\mathbf{h}_{k}\right|^2}{ \sum\limits_{i\in\mathcal{M}_k}\alpha_i\,\gamma_{\mathcal{T}_i}\left|\mathbf{g}_{k}^{\rm H}\,\mathbf{h}_{i}\right|^2 + \norm{\mathbf{g}_{k}}^2},
\end{align}
where $\gamma_{\mathcal{T}_k} $ denotes the effective transmit SNR at $\mathcal{T}_k$ as realized due to the carrier transmission from $\mathcal{R}$, which itself on ignoring the backscattered noise can be defined as:
\begin{equation}\label{eq:SNR-T}
\gamma_{\mathcal{T}_k} \triangleq \frac{1}{\sigma_{\mathrm{w}_\mathcal{R}}^2}\,\sum\limits_{m\in\mathcal{M}}\left|\mathbf{h}_{k}^{\rm T}\mathbf{f}_m\right|^2,\quad\forall k\in\mathcal{M}.
\end{equation} 
Thus, the backscattered-throughput for  $\mathcal{T}_k$ at $\mathcal{R}$ is given by:
\begin{equation}\label{eq:Rk}
\mathrm{R}_k= \log_2\left(1+\gamma_{\mathcal{R}_k}\right),\quad\forall k\in\mathcal{M}.
\end{equation}
{From the above throughput definition which has been extensively used in existing multi-tag BSC investigations, we notice that the key difference from the throughput in conventional networks is the existence of the product or cascaded channels definition and additional BC parameters.}

Lastly, the resulting sum-backscattered-throughput $\mathrm{R_S}$, which is the system-level performance metric as maximized in the current multi-tag monostatic BSC setting, is given by:
\begin{equation}\label{eq:Rs}
\mathrm{R_S}\triangleq\sum_{k\in\mathcal{M}} \,\mathrm{R}_k.
\end{equation} 
Next we use the above sum-throughput definition for carrying out the desired optimization of TRX and BC designs.

\section{Sum Backscattered Throughout Maximization}\label{sec:SRM}
Here we first mathematically formulate the joint optimization problem in Section~\ref{sec:opt-form} and discuss its salient features. Next, after discussing the reasons for non-convexity of the problem, we present the individual optimization schemes for obtaining the  optimal TX precoding, RX beamforming, and BC designs in Sections~\ref{sec:TX},~\ref{sec:RX}, and~\ref{sec:BC}, respectively.

\subsection{Mathematical Optimization Formulation}\label{sec:opt-form}
The joint reader's TRX and tags' BC design to maximize the achievable sum-backscattered-throughput $\mathrm{R}_{\mathrm{S}}$ at $\mathcal{R}$, as defined in \eqref{eq:Rs}, can be mathematically formulated as below:
\begin{eqnarray*}\label{eqOPS}
\begin{aligned} 
	&\mathcal{O}_{\mathrm{S}}: \underset{\left(\mathbf{f}_k,\mathbf{g}_k,\alpha_k\right),\forall k\in\mathcal{M}}{\text{maximize}}\;\mathrm{R_S},\quad\text{subject to}\\
	&
	({\rm C1}): \sum_{k\in\mathcal{M}}\lVert\mathbf{f}_k\rVert^2\le  P_T,\qquad\quad
	({\rm C2}): \lVert\mathbf{g}_k\rVert^2\le 1,\forall k\in\mathcal{M},\\
	&({\rm C3}):\alpha_k\ge\alpha_{\min},\forall k\in\mathcal{M},\quad\;
	({\rm C4}): \alpha_k\le\alpha_{\max},\forall k\in\mathcal{M},
\end{aligned} 
\end{eqnarray*}
where $P_T$ is the available transmit power budget at $\mathcal{R}$, $\alpha_{\min}\ge0$ and $\alpha_{\max}\le1$ respectively the practically-realizable~\cite{QAM-BSC} lower and upper bounds on BC $\alpha_k\in\left(0,1\right)$ for each  tag $\mathcal{T}_k$. {All the computations for obtaining the  jointly-optimal solution of $\mathcal{O}_{\mathrm{S}}$ are performed at $\mathcal{R}$, which then sets its TRX to the optimal one and instructs the tags to set their respective BC accordingly. Here, the battery energy consumption at semi-passive tags in setting their respective BC as per $\mathcal{R}$'s instruction is negligible in comparison to their regular operations~\cite{semi-passive}.}

We notice that although $\mathcal{O}_{\mathrm{S}}$ has convex constraints, it is in general a nonconvex optimization problem because its nonconcave objective includes coupled terms involving the product of optimization variables, i.e., precoders $\mathbf{f}_k$, {combiners} $\mathbf{g}_k$, and BC $\alpha_k$ for each  $\mathcal{T}_k$. Despite the \textit{non-convexity} of joint optimization problem $\mathcal{O}_{\mathrm{S}}$, we here reveal some novel features of the underlying individual optimizations that can yield the \textit{global-optimal} solution for one of them while keeping the other two fixed. In other words, we decouple this problem $\mathcal{O}_{\mathrm{S}}$ into individual optimizations and  then try to solve them separately by exploiting the reduced dimensionality of the underlying problem. We next discuss the individual optimizations, one-by-one, starting with the TX precoder optimization at $\mathcal{R}$ during the downlink carrier transmission.

\subsection{Optimal Transmit Precoding Design at $\mathcal{R}$}\label{sec:TX}
The proposed method for obtaining the optimal TX beamforming vectors $\mathbf{f}_k$ for each $\mathcal{T}_k$ at $\mathcal{R}$ can be divided into two parts. In the first part, we discourse the relationship between  precoder designs for the different tags in the form of Lemma~\ref{lem:prec}. Thereafter proving the concavity of the equivalent semidefinite relaxation (SDR)~\cite{SDR-Multicast} for the precoder design optimization problem, the randomization process~\cite{Phy-Multicast} is used to  ensure desired implicit rank-one constraint on the matrix solution. The implementation steps are provided in the form of Algorithm~\ref{Algo:TBF-Opt}. 
\begin{lemma}\label{lem:prec}
\textit{The optimal precoder designs for the $M$ tags that maximize  the resulting sum-backscattered-throughput $\mathrm{R_S}$ are identical. In other words, $\mathbf{f}_k=\frac{1}{\sqrt{M}}\,\mathbf{f}\in\mathbb{C}^{N\times 1}$ for each $\mathcal{T}_k$.}
\end{lemma}
\begin{IEEEproof} 
{The proof is given in Appendix~\ref{app:prec}. }
\end{IEEEproof}
Lemma~\ref{lem:prec} actually implies that $\mathcal{R}$ transmits with same precoder $\mathbf{f}$ for all the tags, i.e., \textit{multicasting is  optimal TX design}. This is due to the fact that carrier transmission from $\mathcal{R}$ is just to effectively excite (power-up) the tags, and  this excitation can be made most efficient when the TX precoder $\mathbf{f}$ aligns with the strongest eigenmode of the matrix $\sum_{m\in\mathcal{M}}\mathbf{Z}_m$. A similar observation was made in context of the precoder designs for efficient downlink energy transfer in WPCN~\cite{SRM-TVT-WPCN,TGCN-SWIPT}.

{Subsequently,}  with the above result, the sum-backscattered-throughput  $\mathrm{R_S}$ can be rewritten below as a function $\mathrm{R_S^o}\left(\mathbf{f},\mathbf{G},\boldsymbol{\alpha}\right)$  of the common precoding vector $\mathbf{f}$, satisfying $\norm{\mathbf{f}}^2\le P_T$, RX beamforming matrix  $\mathbf{G}\in\mathbb{C}^{N\times M},$ and BC vector  $\boldsymbol{\alpha}\triangleq\left[\alpha_1\;\;\alpha_2\;\;\alpha_3\;\ldots\;\alpha_M\right]^{\rm T}\in\mathbb{R}_{\ge0}^{M\times1}$ for tags:
\begin{eqnarray}\label{eq:Rs1}
\mathrm{R_S^o}\left(\mathbf{f},\mathbf{G},\boldsymbol{\alpha}\right)\triangleq\sum_{k\in\mathcal{M}} \mathrm{R}^\mathrm{o}_k\left(\mathbf{f},\mathbf{g}_k,\boldsymbol{\alpha}\right),
\end{eqnarray}
where backscattered-throughput $\mathrm{R}^\mathrm{o}_k$ as function of $\left(\mathbf{f},\mathbf{g}_k,\boldsymbol{\alpha}\right)$, received SINR $\gamma_{\mathcal{R}_k}^\mathrm{o}$ as a function of $\left(\mathbf{f},\mathbf{g}_k,\boldsymbol{\alpha}\right)$, and transmit SNR $\gamma_{\mathcal{T}_k}^\mathrm{o}$ as a function of $\mathbf{f}$ for $\mathcal{T}_k$ are respectively defined as:
\begin{subequations}\label{eq:Rko}
\begin{gather}
\mathrm{R}^\mathrm{o}_k\left(\mathbf{f},\mathbf{g}_k,\boldsymbol{\alpha}\right)\triangleq \log_2\left(1+\gamma_{\mathcal{R}_k}^\mathrm{o}\right),\quad\forall k\in\mathcal{M},\\
{\gamma_{\mathcal{R}_k}^\mathrm{o}\left(\mathbf{f},\mathbf{g}_k,\boldsymbol{\alpha}\right)\triangleq\frac{\alpha_k\,\gamma_{\mathcal{T}_k}^\mathrm{o} \left|\mathbf{g}_{k}^{\rm H}\,\mathbf{h}_{k}\right|^2}{ \sum\limits_{i\in\mathcal{M}_k}\alpha_i\,\gamma_{\mathcal{T}_i}^\mathrm{o}\left|\mathbf{g}_{k}^{\rm H}\,\mathbf{h}_{i}\right|^2 +\norm{\mathbf{g}_{k}}^2}},\\ \gamma_{\mathcal{T}_k}^\mathrm{o}\left(\mathbf{f}\right)\triangleq \frac{\left|\mathbf{h}_{k}^{\rm T}\mathbf{f}\right|^2}{\sigma_{\mathrm{w}_\mathcal{R}}^2} ,\quad\forall k\in\mathcal{M}.
\end{gather}
\end{subequations} 

However, since $\mathrm{R_S^o}$ is still non-concave in $\mathbf{f}$, we next show that by using an equivalent SDR with matrix definition $\boldsymbol{\mathcal{F}}\triangleq\mathbf{f}\,\mathbf{f}^{\rm H}$  satisfying rank-one constraint, we can resolve this issue.
\begin{lemma}\label{lem:ccF}
\textit{The sum-backscattered-throughput, for a given {combiner} design $\mathbf{G}$ for $\mathcal{R}$ and BC vector  $\boldsymbol{\alpha}$ for the tags, is a concave function of the matrix variable $\boldsymbol{\mathcal{F}}\triangleq\mathbf{f}\,\mathbf{f}^{\rm H}\in\mathbb{C}^{N\times N}$.}
\end{lemma}
\begin{IEEEproof}
	{Please refer to Appendix~\ref{app:ccF} for details.}
\end{IEEEproof}

\begin{algorithm}[!t]
	{\small \caption{Iterative algorithm for precoder $\mathbf{f}$ optimization.}\label{Algo:TBF-Opt}
		\begin{algorithmic}[1]
			\Require 
			\parbox[t]{\dimexpr\linewidth-\algorithmicindent-\algorithmicindent\relax}{Channel vectors $\mathbf{h}_k,\forall k\in\mathcal{M}$, {combiners} $\mathbf{g}_k,\forall k\in\mathcal{M}$, BC  $\boldsymbol{\alpha}$, random samples $K$, {budget $P_T$,} and tolerance $\xi$. \strut}
			\Ensure Global optimal transmit precoding vector $\mathbf{f}_{\rm op}$ 
			\State Set $\mathrm{it}=1,\,\mathbf{f}=\sqrt{P_T}\frac{\sum_{i\in\mathcal{M}}\mathbf{h}_i^*}{\norm{\sum_{i\in\mathcal{M}}\mathbf{h}_i^*}},\,\boldsymbol{\mathcal{F}}^{\rm(it)}=\mathbf{f}\,\mathbf{f}^{\rm H},$ and $\mathrm{R}_{\mathrm{S}}^{(\mathrm{it})}=0$. 
			\Do\Comment{Iteration}
			\State Set  $w_k=\frac{\left|\mathbf{g}_{k}^{\rm H}\,\mathbf{h}_{k}\right|\sqrt{\alpha_k\,{\mathbf{h}_{k}^{\rm T}\boldsymbol{\mathcal{F}}^{\rm(it)}\,\mathbf{h}_{k}^*}}}{\sum\limits_{i\in\mathcal{M}_k}\alpha_i\left|\mathbf{g}_{k}^{\rm H}\,\mathbf{h}_{i}\right|^2\,
				{\mathbf{h}_{i}^{\rm T}\boldsymbol{\mathcal{F}}^{\rm(it)}\,\mathbf{h}_i^*}+\sigma_{\mathrm{w}_{\mathcal{R}}}^2\norm{\mathbf{g}_{k}}^2},\forall k\in\mathcal{M}$.\label{step:wkT}
			\State Set $\mathrm{it}=\mathrm{it}+1$.
			\State 
			\parbox[t]{\dimexpr\linewidth-\algorithmicindent-\algorithmicindent\relax}{Solve the convex problem $\mathcal{O}_{\mathrm{T}_{\rm it}}$ below, satisfying DCP using CVX, and set its global-optimal solution to $\boldsymbol{\mathcal{F}}^{\rm (it)}:$\strut} 
			\begin{eqnarray*}\label{eqOPT-Ti}
				\begin{aligned} 
					\mathcal{O}_{\mathrm{T}_{\rm it}}: \underset{\boldsymbol{\mathcal{F}}}{\text{maximize}}\;& \sum_{k\in\mathcal{M}} \log_2\Bigg[1+ 2\,w_k\left|\mathbf{g}_{k}^{\rm H}\,\mathbf{h}_{k}\right|\sqrt{\alpha_k\,{\mathbf{h}_{k}^{\rm T}\boldsymbol{\mathcal{F}}\,\mathbf{h}_{k}^*}}\\
					& -w_k^2\bigg({ \sum\limits_{i\in\mathcal{M}_k}\alpha_i{\mathbf{h}_{i}^{\rm T}\boldsymbol{\mathcal{F}}\,\mathbf{h}_i^*}\left|\mathbf{g}_{k}^{\rm H}\,\mathbf{h}_{i}\right|^2 + \norm{\mathbf{g}_{k}}^2}\bigg)\Bigg]\\
					&\text{subject to}\quad
					({\rm C5}),\;({\rm C6}).
				\end{aligned} 
			\end{eqnarray*} \label{step:OT}
			\State 	
			\parbox[t]{\dimexpr\linewidth-\algorithmicindent\relax}{Set $\mathrm{R}_{\mathrm{S}}^{(\mathrm{it})}=\sum\limits_{k\in\mathcal{M}} \log_2\bigg[1+ 2\,w_k\left|\mathbf{g}_{k}^{\rm H}\,\mathbf{h}_{k}\right|\sqrt{\alpha_k\,{\mathbf{h}_{k}^{\rm T}\boldsymbol{\mathcal{F}}^{\rm(it)}\,\mathbf{h}_{k}^*}} -w_k^2\bigg({ \sum\limits_{i\in\mathcal{M}_k}\alpha_i\,{\mathbf{h}_{i}^{\rm T}\boldsymbol{\mathcal{F}}^{\rm(it)}\,\mathbf{h}_i^*}\left|\mathbf{g}_{k}^{\rm H}\,\mathbf{h}_{i}\right|^2 + \norm{\mathbf{g}_{k}}^2}\bigg)\bigg].$\strut}
			\doWhile{$\left(\mathrm{R}_{\mathrm{S}}^{(\mathrm{it})}-\mathrm{R}_{\mathrm{S}}^{(\mathrm{it}-1)}\right)\ge\xi$}\Comment{Termination}
			\State Set $\boldsymbol{\mathcal{F}}_{\rm op}=\boldsymbol{\mathcal{F}}^{\rm(it)}$, apply eigenvalue  decomposition to obtain: $\boldsymbol{\mathcal{F}}_{\rm op}=\mathbf{U}\boldsymbol{\Lambda}\mathbf{U}^{\rm H}$, and set $m=1$.\Comment{Randomization}\label{step:RandPs}
			\State Generate $K$ independent unit-variance zero-mean circularly-symmetric complex Gaussian vectors $\mathbf{x_a}_k$  and  $K$ independent uniformly distributed random vectors $\boldsymbol{\theta}_k$   on $\left[0,2\pi\right)$, where both these $K$ vectors are $N\times1$ in size. 
			\State Set $\left[\mathbf{x_b}_k\right]_i=\mathrm{e}^{\left[\boldsymbol{\theta}_k\right]_i},$ where $i=1,2,\ldots,N$ and $k=1,2,\ldots,K.$ 
			\Do 		
			\State Set $\mathrm{R}_{\mathrm{S}}^{\left(1,m\right)}=\mathrm{R_S^o}\left(\mathbf{U}\boldsymbol{\Lambda}^{\frac{1}{2}}\mathbf{x_a}_m,\mathbf{G},\boldsymbol{\alpha}\right)$.
			\State Set $\mathrm{R}_{\mathrm{S}}^{\left(2,m\right)}=\mathrm{R_S^o}\left(\left(\mathrm{diag}\left\lbrace\boldsymbol{\mathcal{F}}_{\rm op}\right\rbrace\odot\mathbf{x_b}_m\right),\mathbf{G},\boldsymbol{\alpha}\right)$.
			\State Set $\mathrm{R}_{\mathrm{S}}^{\left(3,m\right)}=\mathrm{R_S^o}\left(\mathbf{U}\boldsymbol{\Lambda}^{\frac{1}{2}}\mathbf{x_b}_m,\mathbf{G},\boldsymbol{\alpha}\right)$. 
			\State Update $m=m+1$
			\doWhile{$m\le K$} 
			\State Set $\left(i_{\rm op},m_{\rm op}\right)\triangleq\underset{i=\{1,2,3\},\,m=\{1,2,\ldots,K\}}{\mathrm{argmax}}\quad\mathrm{R}_{\mathrm{S}}^{\left(i,m\right)}$.		
			\If{$\left(i_{\rm op}=1\right)$}
			\State Set $\mathbf{f}_{\rm op}=\sqrt{P_T}\,\frac{\mathbf{U}\boldsymbol{\Lambda}^{\frac{1}{2}}\mathbf{x_a}_{m_{\rm op}}}{\norm{\mathbf{U}\boldsymbol{\Lambda}^{\frac{1}{2}}\mathbf{x_a}_{m_{\rm op}}}}$.
			\ElsIf{$\left(i_{\rm op}=2\right)$}
			\State Set $\mathbf{f}_{\rm op}=\sqrt{P_T}\,\frac{\mathrm{diag}\left\lbrace\boldsymbol{\mathcal{F}}^{\rm(it)}\right\rbrace\odot\mathbf{x_b}_{m_{\rm op}}}{\norm{\mathrm{diag}\left\lbrace\boldsymbol{\mathcal{F}}^{\rm(it)}\right\rbrace\odot\mathbf{x_b}_{m_{\rm op}}}}$.
			\Else
			\State Set $\mathbf{f}_{\rm op}=\sqrt{P_T}\,\frac{\mathbf{U}\boldsymbol{\Lambda}^{\frac{1}{2}}\mathbf{x_b}_{m_{\rm op}}}{\norm{\mathbf{U}\boldsymbol{\Lambda}^{\frac{1}{2}}\mathbf{x_b}_{m_{\rm op}}}}$.\label{step:RandPe}
			\EndIf
		\end{algorithmic}
	}
\end{algorithm}

Following Lemmas~\ref{lem:prec} and~\ref{lem:ccF} along with definition \eqref{eq:SINR-F}, the TX beamforming optimization for a given {combiner} and BC design can be formulated as the optimization problem below:
\begin{eqnarray*}\label{eqOPT-Tx}
\begin{aligned} 
	\mathcal{O}_{\mathrm{T}}:\,&\underset{\boldsymbol{\mathcal{F}}}{\text{maximize}}\quad&& \overline{\mathrm{R}}_{\mathrm{S}}=\sum_{k\in\mathcal{M}} \,\log_2\left(1+\overline{\gamma}_{\mathcal{R}_k}\right),\\
	&\text{subject to}\quad&&
	({\rm C5}): \mathrm{Tr}\left(\boldsymbol{\mathcal{F}}\right)\le  P_T,
	\qquad({\rm C6}):\boldsymbol{\mathcal{F}}\succeq0,\\
	& &&({\rm C7}):\mathrm{rank}\left(\boldsymbol{\mathcal{F}}\right)=1,
\end{aligned} 
\end{eqnarray*}  
which on ignoring $({\rm C7})$ is a convex problem with objective function to be maximized being concave in $\boldsymbol{\mathcal{F}}$ (cf. Lemma~\ref{lem:ccF}) and constraints being convex. So, although the optimal solution of $\mathcal{O}_{\mathrm{T}}$ can be obtained using any standard convex optimization toolbox,  like the CVX {\texttt{MATLAB}} package~\cite{cvx}, there lie two challenges. First, the objective function $\overline{\mathrm{R}}_{\mathrm{S}}$ does not satisfy the Disciplined Convex Programming (DCP) rule set for using the CVX toolbox~\cite{cvx,boyd} because each summation includes the ratio of linear functions of $\boldsymbol{\mathcal{F}}$. Second issue is that the optimal $\boldsymbol{\mathcal{F}}$ as obtained after solving the SDR  needs to implicitly satisfy the rank-one constraint $({\rm C7})$.

The first of the above-mentioned issues can be resolved using the recently proposed quadratic transform technique for maximizing the multiple ratio concave-convex linear fractional programming problems~\cite{FP-1}. Further as on ignoring $({\rm C7})$, $\mathcal{O}_{\mathrm{T}}$ is a convex problem, the stationary point as obtained using this quadratic transformation yields the global-optimal solution of $\mathcal{O}_{\mathrm{T}}$. Hence, we can obtain the \textit{global-optimal} $\boldsymbol{\mathcal{F}}$ by solving SDR using CVX toolbox. Thereafter, the second issue can be resolved by deploying the \textit{randomization process}~\cite{Phy-Multicast} to ensure the implicit satisfaction of the rank-one constraint $({\rm C7})$. The detailed algorithmic steps resolving these two issues are summarized in iterative Algorithm~\ref{Algo:TBF-Opt}. It starts with an initial precoding matrix $\boldsymbol{\mathcal{F}}=\mathbf{f}\,\mathbf{f}^{\rm H}$ with $\mathbf{f}=\sqrt{P_T}\frac{\sum_{i\in\mathcal{M}}\mathbf{h}_i^*}{\norm{\sum_{i\in\mathcal{M}}\mathbf{h}_i^*}}$,  which is motivated by the fact that for single-tag case, the respective maximal ratio transmission (MRT) design is optimal~\cite{TSP19}.
Then, after initializing the auxiliary variable  vector  $\boldsymbol{w}\triangleq\left[w_1\;\;w_2\;w_3\;\ldots\;w_M\right]$ as in step~\ref{step:wkT}, we apply the quadratic transformation as suggested in~\cite[Theorem 1]{FP-1}  to each underlying 
SINR term $\overline{\gamma}_{\mathcal{R}_k}$ in~\eqref{eq:SINR-F} and maximize the  corresponding convex reformulation with respect to $\boldsymbol{\mathcal{F}}$, for a given $\boldsymbol{w}$, as denoted by $\mathcal{O}_{\mathrm{T}_{\rm it}}$ in step~\ref{step:OT}. Thereafter, we continue to update  $\boldsymbol{w}$ and optimize $\boldsymbol{\mathcal{F}}$ in an iterative fashion. Since the sum-backscattered-throughput is  concave in $\boldsymbol{\mathcal{F}}$, this sequence  of  convex  optimization problems $\mathcal{O}_{\mathrm{T}_{\rm it}}$ converges  to  a  stationary  point  of $\mathcal{O}_{\mathrm{T}}$, which is also its global-optimal solution,  with nondecreasing  values for the underlying objective  after each iteration. When this improvement in the throughput value reduces below a certain acceptable threshold, the Algorithm~\ref{Algo:TBF-Opt} terminates with the global-optimal precoding matrix $\boldsymbol{\mathcal{F}}_{\rm op}$.  Next, for this precoding solution to satisfy the rank-one constraint $({\rm C7})$ we deploy the randomization process~\cite{SDR-Multicast} as given by steps \ref{step:RandPs} to \ref{step:RandPe} of Algorithm~\ref{Algo:TBF-Opt} which returns the optimal TX precoder $\mathbf{f}_{\rm op}$. The randomization process involves generation of $3K$ set of candidate weight vectors and selecting the one which yields the highest sum-backscattered-throughput among them. Here, we have set $K=10 NM$ samples as mentioned in the results section of \cite{Phy-Multicast} because it maintains a good tradeoff between the solution quality and complexity.

 
\subsection{Receive Beamforming or {Combiner} Design at Reader} \label{sec:RX}
For a given precoder $\mathbf{f}_k$ and BC $\boldsymbol{\alpha}$, the optimal RX beamforming problem is formulated as:
\begin{eqnarray*}\label{eqOPT-Rx}
\begin{aligned} 
	\mathcal{O}_{\mathrm{R}}:&\, \underset{\mathbf{g}_k,\forall k\in\mathcal{M}}{\text{maximize}}\quad \mathrm{R_S},\qquad\text{subject to}\quad
	({\rm C2}).
\end{aligned} 
\end{eqnarray*}
Below we outline a key result defining the optimal RX beamforming or {combiner} design at $\mathcal{R}$.

\begin{lemma}\label{lem:RXBF}
\textit{For a given precoder design $\mathbf{f}_k,\forall k\in\mathcal{M},$ for $\mathcal{R}$ and BC vector  $\boldsymbol{\alpha}$ for the tags, the optimal {combiner} design is characterized by the Wiener or MMSE filter, as defined below:
	\begin{align}\label{eq:MMSE}
	\mathbf{g}_{\mathrm{op}_k}=\frac{\left(\mathbf{I}_N+ \frac{1}{\sigma_{\mathrm{w}_\mathcal{R}}^2} \sum\limits_{i=1}^M
		\alpha_i\left|\mathbf{h}_{i}^{\rm T}\mathbf{f}\right|^2\mathbf{h}_{i} \mathbf{h}_{i}^{\rm H} \right)^{-1}\mathbf{h}_{k}}{\norm{\left(\mathbf{I}_N+ \frac{1}{\sigma_{\mathrm{w}_\mathcal{R}}^2} \sum\limits_{i=1}^M
			\alpha_i\left|\mathbf{h}_{i}^{\rm T}\mathbf{f}\right|^2\mathbf{h}_{i} \mathbf{h}_{i}^{\rm H} \right)^{-1}\mathbf{h}_{k} }}, \forall k\in\mathcal{M}.
	\end{align}}
\end{lemma}
\begin{IEEEproof}
Firstly, from \eqref{eq:SINR} and \eqref{eq:Rk} we notice that $\mathrm{R}_k$ for each  $\mathcal{T}_k$ depends only on its own {combiner} $\mathbf{g}_k$. {Accordingly,} we can maximize the individual rates $\mathrm{R}_k$ or SINRs $\gamma_{\mathcal{R}_k}$ in parallel with respect to $\mathbf{g}_k$, while satisfying their  underlying normalization constraint $({\rm C2})$.  Further, as the $\gamma_{\mathcal{R}_k}$ in \eqref{eq:SINR} can be alternatively represented as a generalized Rayleigh quotient form~\cite[eq. (16)]{SOC-SPMag},  the optimal {combiner}  $\mathbf{g}_{\mathrm{op}_k}$ for each $\mathcal{T}_k$, can be obtained as the generalized eigenvector of the matrix set $\left(\alpha_k\,\gamma_{\mathcal{T}_k}\,\mathbf{h}_{k}\,\mathbf{h}_{k}^{\rm H},\,\sum_{i\in\mathcal{M}_k}\alpha_i\,\gamma_{\mathcal{T}_i}\,\mathbf{h}_{i}\,\mathbf{h}_{i}^{\rm H}+\mathbf{I}_N\right)$ with  largest eigenvalue. Using it along with $({\rm C2})$ and Lemma~\ref{lem:prec}, the optimal {combiner}  in \eqref{eq:MMSE} is obtained.	 
\end{IEEEproof}

\subsection{Backscattering Coefficient (BC) Optimization at Tags}\label{sec:BC}  
Mathematical formulation for this case is presented below:
\begin{eqnarray*}\label{eqOPT-BC}
\begin{aligned} 
	\mathcal{O}_{\mathrm{B}}:&\, \underset{\alpha_k,\forall k\in\mathcal{M}}{\text{maximize}}\quad\mathrm{R_S},\qquad\text{subject to}\quad
	({\rm C3}),\;({\rm C4}).
\end{aligned} 
\end{eqnarray*} 
We would like to mention that although $\mathcal{O}_{\mathrm{B}}$ is a nonconvex problem, it can be solved globally, but in \textit{non-polynomial} time, using the block approximation approach~\cite{block-power}.
Furthermore, even though the sum-backscattered-throughput $\mathrm{R_S}$ defined in \eqref{eq:Rs} is  nonconcave function of  the BC vector $\boldsymbol{\alpha}$, below we present a key property for the   backscattered-throughput $\mathrm{R}_k$ for each $\mathcal{T}_k$, which we have exploited in designing a \textit{computationally-efficient} solution methodology.
\begin{proposition}
\textit{For a given precoder and {combiner} design $\left(\mathbf{f},\mathbf{G}\right)$ for $\mathcal{R}$, both the backscattered SINR
	$\gamma_{\mathcal{R}_k}$ and throughput $\mathrm{R}_k=\log_2\left(1+\gamma_{\mathcal{R}_k}\right)$ for each tag $\mathcal{T}_k$ is pseudolinear in $\boldsymbol{\alpha}$.}
\end{proposition}
\begin{IEEEproof}
As SINR $\gamma_{\mathcal{R}_k}$ involves the ratio of two linear functions, $\alpha_k\gamma_{\mathcal{T}_k} \left|\mathbf{g}_{k}^{\rm H}\,\mathbf{h}_{k}\right|^2$ and ${ \sum_{i\in\mathcal{M}_k}\alpha_i\,\gamma_{\mathcal{T}_i}\left|\mathbf{g}_{k}^{\rm H}\,\mathbf{h}_{i}\right|^2 + \norm{\mathbf{g}_{k}}^2}$ of $\boldsymbol{\alpha}$, using the results of \cite[Tables 5.5 and 5.6]{avriel2010generalized} we note that $\gamma_{\mathcal{R}_k}$ is both pseudoconvex and pseudoconcave in the BC vector $\boldsymbol{\alpha}$. {Now as functions which are both pseudoconvex and pseudoconcave are called pseudolinear~\cite{Baz}, each $\gamma_{\mathcal{R}_k}$ is pseudolinear in $\boldsymbol{\alpha}$.} Further, since the monotonic transformations preserve  pseudolinearity of a function~\cite{avriel2010generalized}, we observe that  throughput $\mathrm{R}_k=\log_2\left(1+\gamma_{\mathcal{R}_k}\right)$ is also pseudolinear in $\boldsymbol{\alpha}$.
\end{IEEEproof}

Here, it is worth noting that since the summation operation does not perverse pseudolinearity~\cite{Baz}, the sum-backscattered-throughput maximization $\mathcal{O}_{\mathrm{B}}$ with respect to $\boldsymbol{\alpha}$ is not a convex problem and hence does not possess global-optimality. {However, we notice that $\mathcal{O}_{\mathrm{B}}$ can be alternatively casted as an optimal power control problem for the sum-rate maximization over the multiple interfering links~\cite[and references therein]{SINR-Power-Opt1}.} For instance, recently in \cite{FP-1} an application of fractional-programming was proposed for efficiently obtaining a stationary point for the nonconvex power control problem over the multiple interfering links. We have used that to yield an efficient low-complexity suboptimal design for BC vector $\boldsymbol{\alpha}$.
\begin{algorithm}[!t]
{\small \caption{Suboptimal BC $\boldsymbol{\alpha}$ optimization algorithm.}\label{Algo:BC-Opt}
	\begin{algorithmic}[1]
		\Require 
		\parbox[t]{\dimexpr\linewidth-\algorithmicindent-\algorithmicindent\relax}{Channel vectors $\mathbf{h}_k,\forall k\in\mathcal{M}$, precoder $\mathbf{f}$, {combiners} $\mathbf{g}_k,\forall k\in\mathcal{M}$, and acceptable tolerance $\xi$.\strut} 
		\Ensure Suboptimal BC  $\boldsymbol{\alpha}_{\rm op}=\left[\alpha_{\mathrm{op}_1}\;\;\alpha_{\mathrm{op}_2}\;\;\alpha_{\mathrm{op}_3}\;\ldots\;\alpha_{\mathrm{op}_M}\right]^{\rm T}$.
		\State Set $\mathrm{it}=1,\,\alpha_k^\mathrm{(it)}=\alpha_{\max},\;\forall k\in\mathcal{M},$ and $\mathrm{R}_{\mathrm{S}}^{(\mathrm{it})}=0$. 
		\Do\Comment{Iteration}
		\State Update  $w_k=\frac{\left|\mathbf{g}_{k}^{\rm H}\,\mathbf{h}_{k}\right|\sqrt{\alpha_k^\mathrm{(it)}\,\gamma_{\mathcal{T}_k}}}{ \sum_{i\in\mathcal{M}_k}\alpha_i^\mathrm{(it)}\,\gamma_{\mathcal{T}_i}\left|\mathbf{g}_{k}^{\rm H}\,\mathbf{h}_{i}\right|^2 + \norm{\mathbf{g}_{k}}^2},\;\forall k\in\mathcal{M}.$\label{step:wk}
		\State Set $\mathrm{it}=\mathrm{it}+1$.
		\State 
		\parbox[t]{\dimexpr\linewidth-\algorithmicindent-\algorithmicindent\relax}{Solve convex problem $\mathcal{O}_{\mathrm{B}_{\rm it}}$ below and set the resulting global-optimal solution to $\left(\alpha_1^\mathrm{(it)},\alpha_2^\mathrm{(it)},\ldots,\alpha_M^\mathrm{(it)}\right):$\strut} 
		\begin{eqnarray*}\label{eqOPT-Bi}
			\begin{aligned} 
				\mathcal{O}_{\mathrm{B}_{\rm it}}: \;&\underset{\alpha_k,\forall k\in\mathcal{M}}{\text{maximize}}\quad \sum_{k\in\mathcal{M}} \log_2\bigg(1+ 2\,w_k\left|\mathbf{g}_{k}^{\rm H}\,\mathbf{h}_{k}\right|\sqrt{\alpha_k\,\gamma_{\mathcal{T}_k}}-
				\\&\qquad\qquad\quad w_k^2\left({ \textstyle\sum_{i\in\mathcal{M}_k}\alpha_i\,\gamma_{\mathcal{T}_i}\left|\mathbf{g}_{k}^{\rm H}\,\mathbf{h}_{i}\right|^2 + \norm{\mathbf{g}_{k}}^2}\right)\bigg),\\
				&\text{subject to}\quad
				({\rm C3}),\;({\rm C4}).
			\end{aligned} 
		\end{eqnarray*} \label{step:OBi}
		\State 
		\parbox[t]{\dimexpr\linewidth-\algorithmicindent-\algorithmicindent\relax}{Set $\mathrm{R}_{\mathrm{S}}^{(\mathrm{it})}=\sum_{k\in\mathcal{M}} \log_2\bigg(1+ 2\,w_k\left|\mathbf{g}_{k}^{\rm H}\,\mathbf{h}_{k}\right|\sqrt{\alpha_k^\mathrm{(it)}\,\gamma_{\mathcal{T}_k}}$\\
			\qquad$-w_k^2\left({ \sum_{i\in\mathcal{M}_k}\alpha_i^\mathrm{(it)}\,\gamma_{\mathcal{T}_i}\left|\mathbf{g}_{k}^{\rm H}\,\mathbf{h}_{i}\right|^2 + \norm{\mathbf{g}_{k}}^2}\right)\bigg).$\strut}
		\doWhile{$\left(\mathrm{R}_{\mathrm{S}}^{(\mathrm{it})}-\mathrm{R}_{\mathrm{S}}^{(\mathrm{it}-1)}\right)\ge\xi$.}\Comment{Termination}
		\State Return $\alpha_{\mathrm{op}_k}=\alpha_k^\mathrm{(it)},\,\forall k\in\mathcal{M}$.
	\end{algorithmic}
}
\end{algorithm}

The detailed algorithmic implementation is outlined in Algorithm~\ref{Algo:BC-Opt}. It starts with an initial BC vector $\boldsymbol{\alpha}$  with all its entries being $\alpha_{\max},$ which is motivated by the fact that for high-SNR regime, the optimal BC is characterized by the full-reflection mode.
Then, after initializing the auxiliary variable  vector  $\boldsymbol{w}$ as in step~\ref{step:wk}, we apply the quadratic transformation as suggested in~\cite[Theorem 1]{FP-1}  to the underlying  each
SINR term and maximize corresponding convex reformulation with respect to $\boldsymbol{\alpha}$, for a given $\boldsymbol{w}$, as denoted by $\mathcal{O}_{\mathrm{B}_{\rm it}}$ in step~\ref{step:OBi}. Thereafter, we continue to update  $\boldsymbol{w}$ and optimize $\boldsymbol{\alpha}$ in an iterative fashion. Since each throughput term is nondecreasing and concave in its respective SINR term, which itself is pseudolinear in  $\boldsymbol{\alpha}$, this sequence  of  convex problems $\mathcal{O}_{\mathrm{B}_{\rm it}}$ converges  to  a  stationary  point  of $\mathcal{O}_{\mathrm{B}}$  with nondecreasing  values for the underlying objective  after each iteration. When this improvement in throughput value reduces below a tolerance $\xi$, the  Algorithm~\ref{Algo:BC-Opt} terminates with a near-optimal BC $\boldsymbol{\alpha}_{\rm op}$.

\section{Proposed Low-Complexity Optimal Designs}\label{sec:Soln} 
{Using the key insights developed for the individually-optimal TX precoding, RX beamforming, and BC vector designs in previous section, now we focus on deriving the jointly-optimal TRX and BC designs by simultaneously solving the original joint optimization problem $\mathcal{O}_{\mathrm{S}}$ (but with $\mathbf{f}_k=  M^{-\frac{1}{2}}\,\mathbf{f}$ for each $\mathcal{T}_k$ based on Lemma~\ref{lem:prec}) in the three optimization variables $\mathbf{f},\mathbf{G},$ and $\boldsymbol{\alpha}$.  We start with presenting novel asymptotically-optimal joint designs for the TRX at $\mathcal{R}$ and BC at tags in both low and high SNR regimes.} In this context, first a joint design for  low-SNR application scenarios is proposed, followed by the other one for the high-SNR regime. These two efficient \textit{low-complexity asymptotically-optimal designs} shed new key design insights on the bounds for  jointly-global-optimal solution. Thereafter, we  conclude by presenting a Nelder--Mead (NM) method~\cite[Ex. 8.51]{Baz} based low-complexity iterative algorithm that does not require the explicit computation of complex derivatives for the objective sum-backscattered-throughput. 
\subsection{Asymptotically-Optimal Design Under High-SNR Regime}\label{sec:high}  
First from Lemma~\ref{lem:RXBF} we revisit that regardless of the precoder and BC design, the optimal {combiner} is characterized by the MMSE filtering defined in~\eqref{eq:MMSE}. Next, we recall that under the high-SNR regime, the ZF-based RX beamforming is known to be a very good approximation for the Wiener or MMSE filter~\cite[eq. (14)]{SOC-SPMag}. So, using the  definition below,
\begin{equation}
{\mathbf{H}}\triangleq\left[{\mathbf{h}}_1\;\;{\mathbf{h}}_2\;\;{\mathbf{h}}_3\;\ldots\;{\mathbf{h}}_M\right]\in\mathbb{C}^{N\times M},
\end{equation} 
the ZF based {combiner} matrix $\mathbf{G}_{\rm Z}\in\mathbb{C}^{N\times M}$ is given by: 
\begin{equation}\label{eq:Gz}
\mathbf{G}_{\rm Z}=\mathbf{H}\,\left(\mathbf{H}^{\rm H}\mathbf{H}\right)^{-1}.
\end{equation} 
As the RX beamforming vector $\mathbf{g}_k$ has to satisfy constraint $\mathrm{(C2)}$, the optimal {combiner} for the high-SNR scenarios, as obtained from $\mathbf{G}_{\rm Z}$ in \eqref{eq:Gz}, is given below:
\begin{equation}
\mathbf{g}_{\mathrm{H}_k}=\frac{\left[\mathbf{G}_{\rm Z}\right]_k}{\norm{\left[\mathbf{G}_{\rm Z}\right]_k}},\quad\forall k\in\mathcal{M},
\end{equation} 
Here, the ZF-based RX beamforming vectors $\mathbf{g}_{\mathrm{H}_k}$ satisfy:
\begin{equation}
\mathbf{g}_{\mathrm{H}_k}^{\rm H}\,\mathbf{h}_{i}=\begin{cases}
0, & \text{$k\neq i$}\\
{\norm{\big[\mathbf{G}_{\rm Z}\big]_k}}^{-1}, & \text{$k= i$,}
\end{cases},\quad\forall k\in\mathcal{M}.
\end{equation}
Thus, with  $\widetilde{\gamma}_{\mathrm{g}_k}\triangleq\frac{1}{\sigma_{\mathrm{w}_{\mathcal{R}}}^2{\norm{\big[\mathbf{G}_{\rm Z}\big]_k}}^2}$, the sum-backscattered-throughput under high-SNR regime where $\mathcal{R}$ employs ZF based {combiner} $\mathbf{G}_{\rm H}\triangleq\left[\mathbf{g}_{\mathrm{H}_1}\;\;\mathbf{g}_{\mathrm{H}_2}\;\ldots\;\mathbf{g}_{\mathrm{H}_M}\right]$, is given by:
\begin{eqnarray}\label{eq:RsH}
\mathrm{R}_{\mathrm{S_H}}\triangleq\mathrm{R_S^o}\left(\mathbf{f},\mathbf{G}_{\rm H},\boldsymbol{\alpha}\right)= {\sum_{k\in\mathcal{M}}}\, \log_2\left(1+ \alpha_k\,\widetilde{\gamma}_{\mathrm{g}_k} \left|\mathbf{h}_{k}^{\rm T}\mathbf{f}\right|^2\right).
\end{eqnarray}

Next revisiting the matrix definition $\boldsymbol{\mathcal{F}}\triangleq\mathbf{f}\,\mathbf{f}^{\rm H}$, the equivalent SDR for jointly optimizing the remaining variables $\boldsymbol{\mathcal{F}}$ and $\boldsymbol{\alpha}$ is formulated below as $\mathcal{O}_{\mathrm{H}}$, which is followed by Lemma~\ref{lem:high-opt} outlining a key result to be used for solving it.
\begin{eqnarray*}\label{eqOPT:H}
\begin{aligned} 
	\mathcal{O}_{\mathrm{H}}:&\, \;\underset{\boldsymbol{\mathcal{F}},\alpha_1,\alpha_2,\ldots,,\alpha_M}{\text{maximize}}\;\; \overline{\mathrm{R}}_{\mathrm{S_H}}\triangleq\sum_{k\in\mathcal{M}} \log_2\left(1+\alpha_k\,\widetilde{\gamma}_{\mathrm{g}_k}{\mathbf{h}_{k}^{\rm T}\boldsymbol{\mathcal{F}}\,\mathbf{h}_{k}^*} \right),\\
	&\quad\;\text{subject to}\quad({\rm C3})\text{ to }({\rm C7}).
\end{aligned} 
\end{eqnarray*}
\begin{lemma}\label{lem:high-opt}
\textit{$\overline{\mathrm{R}}_{\mathrm{S_H}}$ is concave in $\boldsymbol{\mathcal{F}},$ with optimal $\alpha_k$ being equal to  $\alpha_{\max}$ for each $\mathcal{T}_k$.}
\end{lemma}
\begin{IEEEproof}
	{Please refer to Appendix~\ref{app:high-opt} for the proof.}
\end{IEEEproof}

Using Lemma~\ref{lem:high-opt} and ignoring $({\rm C7})$, we notice that  $\mathcal{O}_{\mathrm{H}}$, with $\alpha_k=\alpha_{\max}$ for each tag, is a convex problem in the optimization variable $\boldsymbol{\mathcal{F}}$. Further, since $\mathcal{O}_{\mathrm{H}}$ satisfies the DCP rule, the CVX toolbox can be used to obtain the optimal $\boldsymbol{\mathcal{F}}$, as denoted by $\boldsymbol{\mathcal{F}}_{\mathrm H}$. However, for this precoding solution to satisfy the rank-one constraint $({\rm C7})$ we need to deploy the randomization process, as discussed in Section~\ref{sec:TX} and implemented via steps \ref{step:RandPs} to \ref{step:RandPe} of Algorithm~\ref{Algo:TBF-Opt} while setting $\boldsymbol{\mathcal{F}}_{\rm op}=\boldsymbol{\mathcal{F}}_{\mathrm H}$ in step~\ref{step:RandPs}, to finally get the optimal precoder $\mathbf{f}_{\mathrm H}$.

\begin{remark}\label{rem:high}
\textit{Under high-SNR regime, optimal precoder $\mathbf{f}_{\mathrm H}$ is obtained by solving SDR $\mathcal{O}_{\mathrm{H}}$ with $\alpha_{\mathrm{H}_k}=\alpha_{\max},\forall k\in\mathcal{M},$ followed by randomization process. Whereas, optimal {combiner} follows ZF based design $\mathbf{G}=\mathbf{G}_{\rm H}$ and all the tags are in full-reflection mode, i.e., BC  vector  $\boldsymbol{\alpha}=\boldsymbol{\alpha}_{\rm H}$.  }
\end{remark}
 
\subsection{Novel Joint Design For Low-SNR Applications}\label{sec:low}  
{Under low-SNR regime, we can use the following two  approximations for simplifying $\mathrm{R_S}$:}
\begin{subequations}
\begin{equation}\label{eq:L-I}
\sum\limits_{i\in\mathcal{M}_k}\alpha_i\left|\mathbf{g}_{k}^{\rm H}\,\mathbf{h}_{i}\right|^2\left|\mathbf{h}_{i}^{\rm T}\mathbf{f}\right|^2+{\sigma_{\mathrm{w}_{\mathcal{R}}}^2}\norm{\mathbf{g}_{k}}^2\approx{\sigma_{\mathrm{w}_{\mathcal{R}}}^2}\norm{\mathbf{g}_{k}}^2,
\end{equation}
\begin{align}\label{eq:L-S}
\sum_{k\in\mathcal{M}}&\log_2\left(1+\frac{\alpha_k\, \left|\mathbf{g}_{k}^{\rm H}\,\mathbf{h}_{k}\right|^2\,\left|\mathbf{h}_{k}^{\rm T}\mathbf{f}\right|^2}{{\sigma_{\mathrm{w}_{\mathcal{R}}}^2}\norm{\mathbf{g}_{k}}^2}\right)\nonumber\\
&\approx \frac{1}{\ln\left(2\right)}\sum_{k\in\mathcal{M}}\frac{\alpha_k\,\left|\mathbf{g}_{k}^{\rm H}\,\mathbf{h}_{k}\right|^2\,\left|\mathbf{h}_{k}^{\rm T}\mathbf{f}\right|^2}{{\sigma_{\mathrm{w}_{\mathcal{R}}}^2}\norm{\mathbf{g}_{k}}^2}.
\end{align}
\end{subequations} 
where \eqref{eq:L-I} is owing to the fact that under low-SNR regime, the backscattered signals from all the other tags, causing interference to the tag of interest, is relatively very low in comparison to the received AWGN.  Whereas, \eqref{eq:L-S} is obtained using the approximation $\log_2\left(1+x\right)\approx\frac{x}{\ln\left(2\right)},\forall x\ll1$. Using these properties, the  sum-backscattered-throughput reduces to:
\begin{align}
\mathrm{R}_{\mathrm{S_L}}=&\, \sum_{k\in\mathcal{M}}\frac{\alpha_k \left|\mathbf{g}_{k}^{\rm H}\,\mathbf{h}_{k}\right|^2\,\left|\mathbf{h}_{k}^{\rm T}\mathbf{f}\right|^2}{\ln\left(2\right){\sigma_{\mathrm{w}_{\mathcal{R}}}^2}\norm{\mathbf{g}_{k}}^2}\stackrel{{(a)}}{\le} \frac{\alpha_{\max}\norm{\mathbf{H}^{\rm T}\mathbf{f}}^2}{\ln\left(2\right)\sigma_{\mathrm{w}_{\mathcal{R}}}^2}\nonumber\\
=&\,  \frac{\alpha_{\max}\,\left(\mathbf{f}^{\rm H}\,\mathbf{H}^*\,\mathbf{H}^{\rm T}\mathbf{f}\right)}{\ln\left(2\right)\sigma_{\mathrm{w}_{\mathcal{R}}}^2},
\end{align} 
where ${(a)}$ is based on the individual optimizations of {combiner} and BC vector respectively following MRC and full-reflection mode in this scenario. So, with above as objective $\frac{\alpha_{\max}\,\left(\mathbf{f}^{\rm H}\,\mathbf{H}^*\,\mathbf{H}^{\rm T}\mathbf{f}\right)}{\ln\left(2\right)\sigma_{\mathrm{w}_{\mathcal{R}}}^2}$ and $\mathbf{f}$ as variable, the corresponding maximization problem can be formulated as below:
 
\begin{eqnarray*}\label{eqOPT-TB-L}
\begin{aligned} 
	\mathcal{O}_{\mathrm{TL}}:&\, \underset{\mathbf{f}}{\text{maximize}}\quad  \frac{\alpha_{\max}\,\left(\mathbf{f}^{\rm H}\,\mathbf{H}^*\,\mathbf{H}^{\rm T}\mathbf{f}\right)}{\ln\left(2\right)\sigma_{\mathrm{w}_{\mathcal{R}}}^2},\\
	&\,\text{subject to}\quad
	({\rm C8}): \lVert\mathbf{f}\rVert^2\le  P_T.
\end{aligned} 
\end{eqnarray*}
From $\mathcal{O}_{\mathrm{TL}}$, we notice that the TX precoder design $\mathbf{f}$ at $\mathcal{R}$ that maximizes the sum received power at the tags also eventually yields the maximum  sum-backscattered-throughput from them.	
Thus, the optimal precoder, same for all tags and  called \textit{TX energy beamforming (EB)}, is denoted by: 
\begin{eqnarray}
\mathbf{f}_{\rm L}\triangleq\sqrt{P_T}\;\frac{\mathbf{v}_{\rm max}\left\lbrace\mathbf{H}^*\,{\mathbf{H}}^{\rm T}\right\rbrace}{\norm{\mathbf{v}_{\rm max}\left\lbrace\mathbf{H}^*\,{\mathbf{H}}^{\rm T}\right\rbrace}},
\end{eqnarray}
where $\mathbf{v}_{\rm max}\left\lbrace\mathbf{H}^*\,{\mathbf{H}}^{\rm T}\right\rbrace$ is the right singular vector of the matrix $\mathbf{H}^*\,{\mathbf{H}}^{\rm T}$ that corresponds to its maximum  eigenvalue $\lambda_{\rm max}\left\lbrace\mathbf{H}^*\,{\mathbf{H}}^{\rm T}\right\rbrace$.  So, the total sum received power at tags is:
\begin{equation}\label{eq:RP-SRM}
P_{R}= \mathbf{f}_{\rm L}^{\rm H}\,\mathbf{H}^*\,\mathbf{H}^{\rm T}\mathbf{f}_{\rm L}=  P_T \,\lambda_{\rm max}.
\end{equation} 
On substituting $\mathbf{f}=\mathbf{f}_{\rm L}$ and $\alpha_k=\alpha
_{\rm max}$, for each $\mathcal{T}_k$, in \eqref{eq:MMSE} and using Lemma~\ref{lem:RXBF}, the optimal {combiner} (MMSE based design) at tag $\mathcal{T}_k$ for the low-SNR regime is given by: 
\begin{align}\label{eq:det-L}
\mathbf{g}_{\mathrm{L}_k}\triangleq\frac{\left(\mathbf{I}_N+ \frac{P_T\alpha_{\max}\sum\limits_{i=1}^M
		\left|\mathbf{h}_{i}^{\rm T}\mathbf{v}_{\rm max}\left\lbrace\mathbf{H}^*{\mathbf{H}}^{\rm T}\right\rbrace\right|^2\mathbf{h}_{i}\mathbf{h}_{i}^{\rm H}}{\sigma_{\mathrm{w}_\mathcal{R}}^2} \right)^{-1}\mathbf{h}_{k}}{\norm{\left(\mathbf{I}_N+ \frac{P_T\alpha_{\max}\sum\limits_{i=1}^M
		\left|\mathbf{h}_{i}^{\rm T}\mathbf{v}_{\rm max}\left\lbrace\mathbf{H}^*{\mathbf{H}}^{\rm T}\right\rbrace\right|^2\mathbf{h}_{i}\mathbf{h}_{i}^{\rm H}}{\sigma_{\mathrm{w}_\mathcal{R}}^2} \right)^{-1}\mathbf{h}_{k} }}.
\end{align} 

With precoder and {combiner} designs in low-SNR obtained, next we focus on BC optimization:
\begin{eqnarray*}\label{eqOPT-BC-L}
\begin{aligned} 
	\mathcal{O}_{\mathrm{BL}}:\,& \underset{\alpha_k,\forall k\in\mathcal{M}}{\text{maximize}}\; \sum_{k\in\mathcal{M}} \frac{\left(\frac{1}{\ln\left(2\right)}\right)\alpha_k\,\left|\mathbf{g}_{k}^{\rm H}\,\mathbf{h}_{k}\right|^2 			\left|\mathbf{h}_{k}^{\rm T}\mathbf{f}\right|^2}{\sum\limits_{i\in\mathcal{M}_k}\alpha_i\left|\mathbf{g}_{k}^{\rm H} \mathbf{h}_{i}\right|^2
		\left|\mathbf{h}_{i}^{\rm T}\mathbf{f}\right|^2+{\sigma_{\mathrm{w}_{\mathcal{R}}}^2}\norm{\mathbf{g}_{k}}^2},\\
	&\text{subject to}\quad
	({\rm C3}),({\rm C4}).
\end{aligned} 
\end{eqnarray*} 
Below we present the asymptotically-optimal solution $\boldsymbol{\alpha}_{\rm L}\in\mathbb{R}^{M\times1}$ for $\mathcal{O}_{\mathrm{BL}}$ via Lemma~\ref{lem:BL}.
\begin{lemma}\label{lem:BL}
\textit{For low-SNR settings, where  $\log_2\left(1+\gamma_{\mathcal{R}_k} \right)\approx\frac{\gamma_{\mathcal{R}_k} }{\ln\left(2\right)}$, the optimal BC $\alpha_k$ for each tag $\mathcal{T}_k$, as denoted by $\left[\boldsymbol{\alpha}_{\rm L}\right]_k$, is only characterized either by $\alpha_{\max}$ or $\alpha_{\min}$.}
\end{lemma}
 
\begin{IEEEproof}
Firstly, the result below in \eqref{eq:cvx-SINR}, shows that the sum of SINRs $\gamma_{\mathcal{R}_k} ^{\mathrm{sum}}\triangleq\sum\limits_{k\in\mathcal{M}}\gamma_{\mathcal{R}_k} $ is strictly-convex in $\alpha_k$,
\begin{align}\label{eq:cvx-SINR}
\frac{\partial^2\gamma_{\mathcal{R}_k} ^{\mathrm{sum}}}{\partial\alpha_k^2}&\;=\sum\limits_{i\in\mathcal{M}_k}\frac{2\left(\gamma_{\mathcal{T}_k} \left|\mathbf{g}_{k}^{\rm H}\,\mathbf{h}_{k}\right|^2\right)^2\alpha_i\,\gamma_{\mathcal{T}_i} \left|\mathbf{g}_{i}^{\rm H}\,\mathbf{h}_{i}\right|^2}{\left( \sum_{m\in\mathcal{M}_i} \alpha_m\,\gamma_{\mathcal{T}_m}\left|\mathbf{g}_{i}^{\rm H}\,\mathbf{h}_{m}\right|^2 + \norm{\mathbf{g}_{i}}^2\right)^3}\nonumber\\
&\;>0.
\end{align}
Next since we aim to maximize the scaled $\gamma_{\mathcal{R}_k} ^{\mathrm{sum}}$ in $\mathcal{O}_{\mathrm{BL}}$  and the maximum value of a convex function lies at the corner points of its underlying variable, we conclude that the optimal value of each $\alpha_k$ is set to either one out of $\alpha_{\max}$ or $\alpha_{\min}$.
\end{IEEEproof}
\begin{remark}\label{rem:low}
\textit{Under low-SNR regime, precoding $\mathbf{f}_{\rm L}$ reduces to TX-EB and {combiner} design follows MMSE filtering (cf. \eqref{eq:det-L}). Whereas, BC optimization reduces to a low-complexity binary decision-making process, in which just $2^M-1$ possible candidates need to be checked for $\boldsymbol{\alpha}$ to eventually select the best $\boldsymbol{\alpha}_{\rm L}$ among them in terms of the sum-throughput.}
\end{remark}
From Remarks~\ref{rem:high} and~\ref{rem:low}, we notice that  for the asymptotic cases, the  optimal RX and BC designs are available in closed-form in terms of the TX precoder, where the latter can be numerically computed efficiently using either SDR followed by randomization, or  eigenvalue-decomposition.

\subsection{Low-Complexity Algorithm for Jointly-Suboptimal Design}\label{sec:Alg} 
With the two asymptotically-optimal designs as obtained in the two previous subsections, now we develop a low-complexity iterative algorithm that uses them to present an efficient suboptimal joint design. Since, the low and high-SNR regimes form the two extremes (in terms of SNR boundaries) from a geometrical viewpoint, the optimal TX beamforming or precoding vector $\mathbf{f}$ for any arbitrary (or finite) SNR needs to balance between these two extremes. 
\begin{remark}
\textit{In other words. the optimal TX precoding vector is based on the direction that trade-offs between the following two contradictory objectives of:
	\begin{itemize}
		\item maximizing the sum-received RF power among the tags  by implementing TX-EB~\cite{TGCN-SWIPT} during the downlink carrier transmission with the precoder set as $\mathbf{f}_{\mathrm{L}}$, and
		\item  balancing between the individual MRT direction for each tag as in case of single-group multicasting based downlink transmission~\cite{Phy-Multicast} and  setting the precoder as $\mathbf{f}_{\mathrm{H}}$.
	\end{itemize} }
\end{remark} 

Capitalizing on this insight, we propose the following weighted TX beamforming direction: 
\begin{equation}\label{eq:wTXB}
\mathbf{f}_{\rm w} \triangleq \frac{\boldsymbol{\mathrm{w}}\odot\mathbf{f}_{\mathrm{L}}+\left(\mathbf{1}_{N\times1}-\boldsymbol{\mathrm{w}}\right)\odot\mathbf{f}_{\mathrm{H}}}{\norm{\boldsymbol{\mathrm{w}}\odot\mathbf{f}_{\mathrm{L}}+\left(\mathbf{1}_{N\times1}-\boldsymbol{\mathrm{w}}\right)\odot\mathbf{f}_{\mathrm{H}}}},
\end{equation}
where $\boldsymbol{\mathrm{w}}\triangleq\left[\mathrm{w}_1\;\;\mathrm{w}_2\;\mathrm{w}_3\;\ldots\;\mathrm{w}_N\right]^{\rm T}\in\left[0,1\right]^{N\times1}$ represents the relative weight between  asymptotically-optimal TX beamforming direction $\mathbf{f}_{\mathrm{L}}$ in low-SNR regime and the corresponding direction $\mathbf{f}_{\mathrm{H}}$ for high-SNR scenarios. To further reduce the computational complexity of proposed iterative algorithm, we use an uniform weight allocation scheme where $\boldsymbol{\mathrm{w}}=\mathrm{w}_0\,\mathbf{1}_{N\times1}$, and thus, the weighted TX beamforming direction in  \eqref{eq:wTXB} reduces to the precoding vector ${\mathbf{f}}_{\rm w_0}$ defined below: 
\begin{equation}\label{eq:wTXB0}
{\mathbf{f}}_{\rm w_0} \triangleq \frac{\mathrm{w}_0 \,\mathbf{f}_{\mathrm{L}}+\left(1-\mathrm{w}_0\right)\mathbf{f}_{\mathrm{H}}}{\norm{\mathrm{w}_0\,\mathbf{f}_{\mathrm{L}}+\left(1-\mathrm{w}_0\right)\mathbf{f}_{\mathrm{H}}}}.
\end{equation}
{We vary this common  weight $\mathrm{w}_0$  in $K_0$ discrete steps ranging from $0$ to $1$, and thus the resulting weights are $\left\lbrace0,\frac{1}{K_0-1},\frac{2}{K_0-1},\ldots,\frac{K_0-2}{K_0-1},1\right\rbrace$. Here $K_0$ is selected as per the desired  solution-quality versus computational-complexity tradeoff. To compute the optimal $\mathrm{w}_0$ yielding the maximum $\mathrm{R_S}$, one needs to evaluate $\mathrm{R_S}$ for all the $K_0$ weights and then select the best among them.}

{We use ${\mathbf{f}}_{\rm w_0}$ and $\boldsymbol{\alpha}_{\rm H}$ as the starting point for the NM method and then try to maximize the sum-backscattered-throughput by jointly-optimizing $\mathbf{f}$ and $\boldsymbol{\alpha},$ while setting $\mathbf{G}$ based on the MMSE filtering design (cf. \eqref{eq:MMSE}) as their function.} The detailed steps are outlined in Algorithm~\ref{Algo:Opt}. The key merits of using NM method, not involving the calculation of derivatives (or  gradients) which can be computationally very  expensive due to the involvement of matrix-inverse operations in the MMSE-based optimal {combiner} definition, is a low-complexity algorithm inbuilt in most conventional solvers like {\texttt{MATLAB}}. {The NM method is iteratively called $K_0$ times for different starting points in accordance to the $K_0$ weights-based ${\mathbf{f}}_{\rm w_0}$ definition in \eqref{eq:wTXB0} and the one yielding highest sum-backscattered-throughput is selected to yield the proposed jointly-optimal precoder $\mathbf{f}_{\rm J}$ and BC $\boldsymbol{\alpha}_{\rm J}$ design, which eventually are used to obtain the {combiner} design $\mathbf{G}_{\rm J}$ by respectively substituting them in place of  $\mathbf{f}$ and $\boldsymbol{\alpha}$ in~\eqref{eq:MMSE}.} Note that Algorithm~\ref{Algo:Opt} returns a suboptimal joint design yielding higher sum-backscattered-throughput than both of the two asymptotically-optimal joint designs, and the number of NM-method restarts or iterations $K_0$ needs to be judiciously selected based on the desired performance quality and acceptable complexity in achieving that. Further, in Section~\ref{sec:res} we have   verified the fast convergence of Algorithm~\ref{Algo:Opt} via Fig.~\ref{fig:valid}(b).

\begin{algorithm}[!t]
	{\small \caption{Iterative NM Method Based Joint Optimization.}\label{Algo:Opt}
		\begin{algorithmic}[1]
			\Require Channel vectors $\mathbf{h}_k,\forall k\in\mathcal{M},$ and number of iterations $K_0$.
			\Ensure 
			\parbox[t]{\dimexpr\linewidth-\algorithmicindent-\algorithmicindent\relax}{Jointly-optimal precoder $\mathbf{f}_{\rm J}$, {combiners} $\mathbf{g}_{\rm J_k},\forall k\in\mathcal{M}$, BC vector $\boldsymbol{\alpha}_{\rm J}$, and  budget $P_T$  along with throughput $\mathrm{R}_{\mathrm{S}_{\rm J}}$.\strut}
			\State Define  $\widetilde{\mathbf{G}}\left(\mathbf{f},\boldsymbol{\alpha}\right)$ as the matrix with its $M$ columns being:
			\begin{eqnarray}\nonumber
			 &\widetilde{\mathbf{g}}_{k}\left(\mathbf{f},\boldsymbol{\alpha}\right)=\frac{\left(\mathbf{I}_N+ \frac{1}{\sigma_{\mathrm{w}_\mathcal{R}}^2}\,\sum\limits_{i=1}^M
			 	\alpha_i\left|\mathbf{h}_{i}^{\rm T}\mathbf{f}\right|^2\mathbf{h}_{i}\,\mathbf{h}_{i}^{\rm H} \right)^{-1}\mathbf{h}_{k}}{\norm{\left(\mathbf{I}_N+ \frac{1}{\sigma_{\mathrm{w}_\mathcal{R}}^2}\,\sum\limits_{i=1}^M
			 		\alpha_i\left|\mathbf{h}_{i}^{\rm T}\mathbf{f}\right|^2\mathbf{h}_{i}\,\mathbf{h}_{i}^{\rm H} \right)^{-1}\mathbf{h}_{k} }},\;\forall k\in\mathcal{M}.
			\end{eqnarray}
			\State Obtain $\mathbf{f}_{\rm H}$ and $\boldsymbol{\alpha}_{\rm H}$ respectively as the precoder and BC vector designs for the high-SNR regime.
			\State Set $\mathrm{it}=0,\,\mathrm{R}_{\mathrm{S}}^{\max}=0,$ and obtain $\mathbf{f}_{\rm L}$  as the precoder for the low-SNR regime.\Comment{Initialization}
			\Do\Comment{Iteration}
			\State Update $\mathrm{it}=\mathrm{it}+1$ and $\mathrm{w}_0=\frac{\mathrm{it}-1}{K_0-1}$. 
			\State Update ${\mathbf{f}}_{\rm w_0}=\sqrt{P_T}\,\frac{\mathrm{w}_0\,\mathbf{f}_{\rm L}+\left(1-\mathrm{w}_0\right)\,\mathbf{f}_{\rm H}}{\norm{{\mathrm{w}_0\,\mathbf{f}_{\rm L}+\left(1-\mathrm{w}_0\right)\,\mathbf{f}_{\rm H}}}}$. 
			\State 
			\parbox[t]{\dimexpr\linewidth-\algorithmicindent\relax}{Apply NM method with $\left({\mathbf{f}}_{\rm w_0},\,\boldsymbol{\alpha}_{\rm H}\right)$ as  starting point for jointly maximizing  $\mathrm{R_S^o}\left(\mathbf{f},\widetilde{\mathbf{G}}\left(\mathbf{f},\boldsymbol{\alpha}\right),\boldsymbol{\alpha}\right)$ in $\left(\mathbf{f},\boldsymbol{\alpha}\right)$.\strut}\label{stp:NM} 
			\State 
			\parbox[t]{\dimexpr\linewidth-\algorithmicindent\relax}{Set the resulting joint optimal precoding and BC vector solution in step~\ref{stp:NM}  to $\left(\mathbf{f}^{(\mathrm{it})},\boldsymbol{\alpha}^{(\mathrm{it})}\right)$.\strut} 
			\State 
			\parbox[t]{\dimexpr\linewidth-\algorithmicindent\relax}{Set $\mathbf{G}^{(\mathrm{it})}$ as the matrix with columns $\widetilde{\mathbf{g}}_{k}\left(\mathbf{f}^{(\mathrm{it})},\boldsymbol{\alpha}^{(\mathrm{it})}\right)$ and $\;\mathrm{R}_{\mathrm{S}}^{(\mathrm{it})}=\mathrm{R_S^o}\left(\mathbf{f}^{(\mathrm{it})},\mathbf{G}^{(\mathrm{it})},\boldsymbol{\alpha}^{(\mathrm{it})}\right).$\strut}
			\If{$\left(\mathrm{R}_{\mathrm{S}}^{(\mathrm{it})}\ge\mathrm{R}_{\mathrm{S}}^{\max}\right)$}
			\State 
			\parbox[t]{\dimexpr\linewidth-\algorithmicindent-\algorithmicindent\relax}{Set $\mathrm{R}_{\mathrm{S}}^{\max}=\mathrm{R}_{\mathrm{S}}^{(\mathrm{it})},\,\mathrm{R}_{\mathrm{S}_{\rm J}}=\mathrm{R}_{\mathrm{S}}^{(\mathrm{it})}$,\, $\mathbf{f}_{\rm J}=\mathbf{f}^{(\mathrm{it})},$\,  $\mathbf{G}_{\rm J}=\mathbf{G}^{(\mathrm{it})}$ with columns $\mathbf{g}_{\rm J_k},\forall k\in\mathcal{M}$,\, and  $\boldsymbol{\alpha}_{\rm J}=\boldsymbol{\alpha}^{(\mathrm{it})}$. \strut}
			\EndIf
			\doWhile{$\left(\mathrm{it}\le K_0\right)$}\Comment{Termination}
		\end{algorithmic}
	} 
\end{algorithm}

\begin{remark}\label{rem:TRX}
	\textit{If due to the noncooperation of tags in the optimization process, one wishes to obtain the optimal TRX design $\left(\mathbf{f},\mathbf{G}\right)$ using Algorithm~\ref{Algo:Opt} with fixed BC vector as $\boldsymbol{\alpha}_{\rm H}$ (i.e., all the tags in full-reflection mode), then we just need to modify step~\ref{stp:NM} as: ``Apply NM method with ${\mathbf{f}}_{\rm w_0}$ as  starting point for maximizing  $\mathrm{R_S^o}\left(\mathbf{f},\widetilde{\mathbf{G}}\left(\mathbf{f},\boldsymbol{\alpha}_{\rm H}\right),\boldsymbol{\alpha}_{\rm H}\right)$ in $\mathbf{f}$." Rest of the steps in Algorithm~\ref{Algo:Opt} remain the same, while replacing each $\boldsymbol{\alpha}^{(\mathrm{it})}$ with  $\boldsymbol{\alpha}_{\rm H},\forall\, \mathrm{it}\in\left[1,K_0\right]$. We denote this resulting optimal TRX design at $\mathcal{R}$ for fixed $\boldsymbol{\alpha}=\boldsymbol{\alpha}_{\rm H}$ by $\left(\mathbf{f}_{\rm J{\alpha_H}},\mathbf{G}_{\rm J{\alpha_H}}\right)$.}
\end{remark}  

\section{Design Utilities and Research Extensions}\label{sec:dis} 
In this section we corroborate the practical utility of the proposed TRX and BC designs by showing how they can be used to address the requirements of other BSC models, with or without perfect CSI availability. We also include brief discussion on the extension of these results to the multiantenna tag based BSC and for meeting the requirements of the WPCN systems. {Some of the claims from these discussions will also be supported via simulation results in Section~\ref{sec:res}.}

\subsection{Other Backscatter Communication Settings}\label{sec:other-BSC}
\subsubsection{Nonreciprocal-Monostatic, Bi-static, and Ambient BSC Models}\label{sec:nonReciprocal}
Though the optimal designs presented in this work are dedicated to the monostatic BSC settings with  reciprocal $\mathcal{T}_k$-to-$\mathcal{R}$ channels, these results can be easily extended to the nonreciprocal-monostatic or bi-static BSC systems where the  $\mathcal{T}_k$-to-$\mathcal{R}$ and $\mathcal{R}$-to-$\mathcal{T}_k$ channels are different and are respectively denoted by the forward $\mathbf{h}_{F_k}\in\mathbb{C}^{N\times1}$ (instead of $\mathbf{h}_k$) and backward $\mathbf{h}_{B_k}\in\mathbb{C}^{1\times N}$ (instead of $\mathbf{h}_k^{\rm T}$)  channel vectors for each $\mathcal{T}_k$. {Therefore,} the same results as proposed in Sections~\ref{sec:SRM} and~\ref{sec:Soln} will hold good for the nonreciprocal-monostatic or bi-static BSC settings, but with $\mathbf{g}_{k}^{\rm H}\,\mathbf{h}_{i}$ and $\mathbf{h}_{i}^{\rm T}\mathbf{f}$ being respectively replaced by $\mathbf{g}_{k}^{\rm H}\,\mathbf{h}_{F_i}$ and $\mathbf{h}_{B_i}\mathbf{f},$ $\forall i,k\in\mathcal{M.}$ 

In contrast to the monostatic and bi-static settings, for the ambient BSC scenarios, we can only design the {combiner} and BC  as  carrier transmission is from an uncontrollable (ambient) source. So, following Lemma~\ref{lem:RXBF}, the optimal {combiner} at $\mathcal{R}$  follows MMSE filtering  design. Whereas, the methodology for finding  BC design for  each of these three settings mentioned here is exactly same as that for  reciprocal-monostatic BSC setting investigated in Sections~\ref{sec:BC} and~\ref{sec:Soln}.    

\subsubsection{Non-Availability of Perfect CSI at $\mathcal{R}$}\label{sec:imperfect-CSI}
In this work with the aim of investigating the maximum achievable throughput performance gains due to a large antenna array at $\mathcal{R}$, we focused on the joint optimal TRX and BC design while assuming the perfect CSI availability for the reciprocal backscattered channel at $\mathcal{R}$. However, in practice perfect CSI is not available and we need to design the TRX and BC based on the estimated CSI. Also, CE is more challenging in BSC systems~\cite{New-SP-Mag2}  because the tags do not have their own radio circuitry for processing incoming signals or transmitting uplink pilots to aid in CE.

\paragraph{Obtaining Channel Estimates in Multi-tag BSC Settings}\label{sec:CE}

{Recently in~\cite{TSP19}, a least-squares-estimator (LSE) for the backscattered channel was proposed for the reciprocal monostatic BSC system with full-duplex multiantenna reader and single-antenna tag. Using this LSE along with the tag-switching (binary BC setting, cf. Remark~\ref{rem:low}) based pilot-signal backscattering from the tags as proposed in~\cite{BSC-WET-MIMO}, where only one tag is active during a sub-phase of the CE phase, we can obtain the estimate for each $\mathcal{T}_k$-to-$\mathcal{R}$ channel as $\widehat{\mathbf{h}}_k$. Basically, once the excitation energy is received due to $\mathcal{R}$'s carrier transmission, the tags one-by-one go into a silent period by setting their respective BC to a minimum value, say $\alpha_{\min}$. This tags cooperation  can be seen like the orthogonal preamble sequences  known  at $\mathcal{R}$, which eventually help it to estimate the underlying BSC channels~\cite{BackFi,TSP19,BSC-WET-MIMO}. These estimates are then used for TRX designing and data detection at  multiantenna $\mathcal{R}$. This completes the CE phase. Thereafter, the actual backscattered data  transmission phase starts where the tags then sends their data payload by modulating the signal from $\mathcal{R}$.}  Then, following the discussions on utilization of estimated CSI for TRX designs in~\cite{massive-MIMO,TSP19,BSC-WET-MIMO}, we note that the joint TRX-BC design under non-availability of perfect CSI can be obtained in the same way as discussed in Section~\ref{sec:Soln}, but with each $\mathbf{h}_k$ respectively replaced by its estimate  $\widehat{\mathbf{h}}_k$, because $\mathcal{R}$ treats the estimated channel as the true one.

\paragraph{Robustness Against Imperfect CSI Knowledge}\label{sec:robust}
Now if $\mathcal{R}$ has imperfect knowledge of the instantaneous  realization for BSC channel vector  $\mathbf{h}_k$ for each $\mathcal{T}_k$, then the underlying CSI $\widetilde{\mathbf{h}}_k$ can be modeled using the generic Gauss-Markov formulation as given below~\cite[eq. (13)]{imperfect-CSI} 
\begin{eqnarray}
\widetilde{\mathbf{h}}_k=\sqrt{1-\eta^2}\;\mathbf{h}_k+\eta\,\sqrt{\beta_k}\,\mathbf{z}_k,\quad\forall k\in\mathcal{M},
\end{eqnarray} 
where
vector $\mathbf{z}_k\sim\mathbb{C} \mathbb{N}\left(\mathbf{0}_{N\times 1},\mathbf{I}_N\right)$
accounts for CE errors independent of   $\mathbf{h}_k$ and the scaling parameter $\eta\in\left[0,1\right]$ indicates the quality of the instantaneous CSI. So, $\eta=0$ corresponds to perfect 
CSI case and $\eta=1$ to having only statistical CSI. Hence, when only imperfect CSI is available at $\mathcal{R}$,  we replace each true channel  $\mathbf{h}_k$  by its respective estimate  $\widetilde{\mathbf{h}}_k$. We have verified the impact of inaccuracy parameter $\eta$ on the optimized performance in Sections~\ref{sec:Comp-Semi} and~\ref{sec:bench}.

\subsubsection{Multiantenna Single-Tag Setup}\label{sec:multiantenna-tag} 
This work focuses on the TRX design for serving multiple single-antenna tags. Now we would like to give some insights on the reader's TRX design for serving a single tag $\mathcal{T}_0$ with $M_0$ antenna elements, such that the underlying $\mathcal{T}_0$-to-$\mathcal{R}$ MIMO channel is represented by $\mathbf{H}_0\in\mathbb{C}^{N\times M_0}$. Thus, denoting the $M_0$-element BC vector, one for each antenna at $\mathcal{T}_0$, by $\boldsymbol{\alpha}_0\in\mathbb{R}_{\ge0}^{M_0\times1}$, and following the derivations in Sections~\ref{sec:tag} and~\ref{sec:BSC-T}, the resulting backscattered SNR at $\mathcal{R}$ is given by:
\begin{align}
\gamma_0\triangleq&\,\frac{\left|\mathbf{g}^{\rm H}\,{\mathbf{H}_0}\,\left(\mathrm{diag}\left\lbrace\boldsymbol{\alpha}_0\right\rbrace\right)^{1/2}\,{\mathbf{H}_0}^{\rm T}\,\mathbf{f}\right|^2}{{\sigma_{\mathrm{w}_{\mathcal{R}}}^2}\norm{\mathbf{g}}^2}\nonumber\\
\stackrel{{(b)}}{=}&\,\left(\frac{\alpha_{\max}}{\sigma_{\mathrm{w}_{\mathcal{R}}}^2}\right)\frac{\mathbf{g}^{\rm H}\,{\mathbf{H}_0}\,{\mathbf{H}_0}^{\rm T}\,\mathbf{f}\,\mathbf{f}^{\rm H}\,{\mathbf{H}_0}^*\,{\mathbf{H}_0}^{\rm H}\,\mathbf{g}}{\mathbf{g}^{\rm H}\,\mathbf{g}},
\end{align}
where ${(b)}$ is obtained by noting that for a single tag setup, maximum SNR is achieved under the full-reflection mode, i.e., $\boldsymbol{\alpha}_0\triangleq\alpha_{\max}\mathbf{1}_{M\times1}$.
Next using \cite[eq. (31)]{opt-RX-Mat}, we note that the optimal decoder $\mathbf{g}=\mathbf{g}_0\in\mathbb{C}^{N\times 1}$ for the single user case, following the MRC design with effective channel as  ${\mathbf{H}_0}\,{\mathbf{H}_0}^{\rm T}\,\mathbf{f}$, that maximizes backscattered SNR $\gamma_0$ is given by  $\mathbf{g}_0\triangleq\frac{{\mathbf{H}_0}\,{\mathbf{H}_0}^{\rm T}\,\mathbf{f}}{\norm{{\mathbf{H}_0}\,{\mathbf{H}_0}^{\rm T}\,\mathbf{f}}}.$ 
So, with optimal BC and {combiner} designs $\left(\boldsymbol{\alpha}_0,\mathbf{g}_0\right)$, $\gamma_0$  as  function of precoder $\mathbf{f}$ reduces to:
\begin{align}
{\gamma}_0=&\,\frac{\alpha_{\max}}{\sigma_{\mathrm{w}_{\mathcal{R}}}^2}\frac{{\norm{{\mathbf{H}_0}\,{\mathbf{H}_0}^{\rm T}\,\mathbf{f}}^4}}{\norm{{\mathbf{H}_0}\,{\mathbf{H}_0}^{\rm T}\,\mathbf{f}}^2}=\frac{\alpha_{\max}}{\sigma_{\mathrm{w}_{\mathcal{R}}}^2}{\norm{{\mathbf{H}_0}\,{\mathbf{H}_0}^{\rm T}\,\mathbf{f}}^2}\nonumber\\
=&\,\frac{\alpha_{\max}}{\sigma_{\mathrm{w}_{\mathcal{R}}}^2}\,\left(\mathbf{f}^{\rm H}\,{\mathbf{H}_0}^*\,{\mathbf{H}_0}^{\rm H} \,{\mathbf{H}_0}\,{\mathbf{H}_0}^{\rm T}\,\mathbf{f}\right).
\end{align}
It is well known from \cite{TGCN-SWIPT}, that the  maximum  SNR during the downlink MIMO transmission is achieved by implementing the TX-EB at $\mathcal{R}$. This EB based precoder is characterized by the strongest eigenmode of the
matrix $\widetilde{\mathbf{H}}_0\triangleq{\mathbf{H}_0}^*\,{\mathbf{H}_0}^{\rm H}\,{\mathbf{H}_0}\,{\mathbf{H}_0}^{\rm T} $. Hence, the optimal precoder is given by $\mathbf{f}_0\triangleq \sqrt{P_T}\,{\mathbf{v}}_{\max}\left\lbrace\widetilde{\mathbf{H}}_0\right\rbrace$, where ${\mathbf{v}}_{\max}\left\lbrace\widetilde{\mathbf{H}}_0\right\rbrace$ is the singular vector corresponding to the maximum eigenvalue ${\lambda}_{\max}\left\lbrace\widetilde{\mathbf{H}}_0\right\rbrace$ of matrix $\widetilde{\mathbf{H}}_0$. Thus, optimum backscattered-throughput is given by ${\mathrm{R}}_{\rm S_0}=\log_2\left(1+{\alpha_{\max}\,P_T\,{\lambda}_{\max}\left\lbrace\widetilde{\mathbf{H}}_0\right\rbrace}\left({\sigma_{\mathrm{w}_{\mathcal{R}}}^2}\right)^{-1}\right)$. 
\begin{remark}
	\textit{From the results in this subsection and the ones in Section~\ref{sec:low}, we would like to draw attention on a key observation that under low-SNR regime, the joint TRX and BC design for the single-antenna multi-tag setup is similar to those in the multiantenna single-tag setting.}
\end{remark}
{Lastly, the generic multiple multiantenna tag setting, which is not studied here and also not much in the existing art due to practical limitations of BSC, will be requiring  a totally new and dedicated investigation. It is one of our future research directions for building upon this work.}

\subsection{Wireless Powered Communication Systems}\label{sec:WPCN}
There is a striking similarity between WPCN~\cite{WPCN-MIMO-CTM,SRM-TVT-WPCN,WPCN-MIMO-SRM} and BSC systems because the downlink energy transfer phase in WPCN to power-up the RF energy harvesting (EH) users has similar objective like the reader's carrier transmission to tags for exciting them. This relation for the downlink transmission  leads to a very similar throughput expression for the two systems as can be noted from \cite[eq. (4)]{WPCN-MIMO-SRM} and \eqref{eq:Rs1}, respectively. {Basically, the main difference between the throughput expressions for these two set-ups and other multi-user settings with multiantenna access-point~\cite{SOC-SPMag,massive-MIMO,TxBroadC1,TxBroadC2} is that the precoder terms in the throughput expression (defined by \eqref{eq:Rk} or \eqref{eq:Rko}) for each tag or user are same in BSC or WPCN settings. In contrast, for conventional multi-user set-ups, this throughput expression~\cite[eq. (2)]{SOC-SPMag},~\cite[eq. (3)]{TxBroadC2} for each user is different because for any  user $\mathcal{T}_k$, its useful signal term contains only its own precoding vector $\mathbf{f}_k$, with all other remaining precoders $\mathbf{f}_i,\forall i\in\mathcal{M}_k,$ contributing to the interference term. Hence, the existing TX precoder designs, as proposed for optimal multi-user TX beamforming by exploiting Rayleigh quotient forms~\cite{SOC-SPMag,massive-MIMO}, or for  ergodic sum-rate maximization in broadcast settings by solving underlying eigenvalue problems~\cite{TxBroadC1,TxBroadC2}, cannot be used  for BSC.} 

{However, due  to the above-mentioned similarity between WPCN and BSC settings, the proposed precoder and {combiner} designs can be  applied for the sum-rate-maximization in WPCN, with multiantenna hybrid access point (HAP) and multiple single-antenna EH users, as investigated in~\cite{SRM-TVT-WPCN,WPCN-MIMO-SRM}. Here, it is worth noting that in contrast to~\cite{SRM-TVT-WPCN,WPCN-MIMO-SRM}, where suboptimal TRX designs were proposed for the multiantenna HAP, the optimal solutions in this work outperform them as shown via numerical results in Section~\ref{sec:bench}. The main reason behind the performance enhancement of our proposed TRX designs over the existing ones is the \textit{global-optimality} of individual designs and \textit{asymptotic-optimality} of the low-complexity-suboptimal joint-ones, as discussed in Sections~\ref{sec:SRM} and~\ref{sec:Soln}, respectively. Also, the TX precoder in~\cite{WPCN-MIMO-SRM} was not even individually global-optimal, in contrast to ours,  because the  reformulated problem~\cite[eq. (9)]{WPCN-MIMO-SRM} was not equivalent to the original one~\cite[eq. (8)]{WPCN-MIMO-SRM} leading to the performance gap.}

\begin{figure*}[!t]
	\centering  
	\subfigure[Verifying the quality of proposed low-complexity joint designs.]{{\includegraphics[height=1.5in]{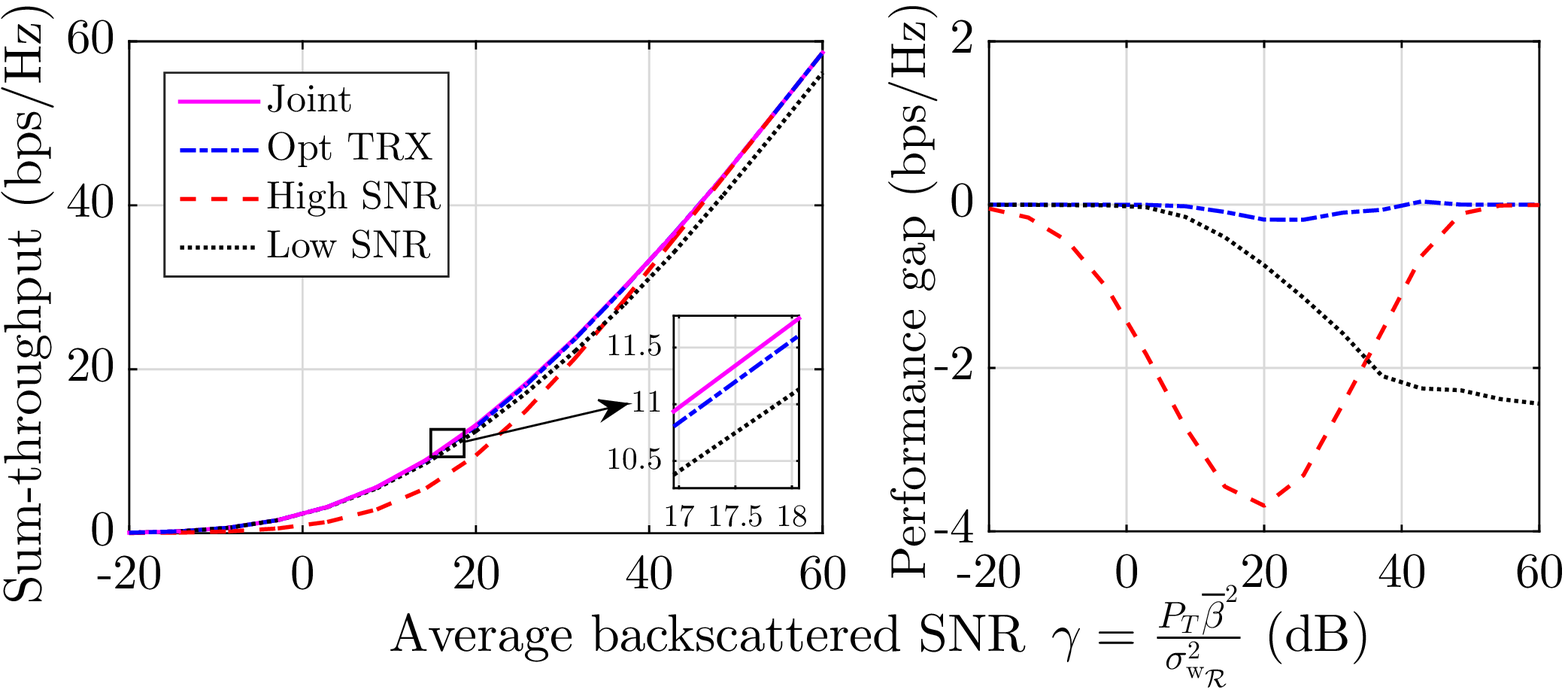} }} \quad  
	\subfigure[Verifying the fast convergence of Algorithm~\ref{Algo:Opt}.]{{\includegraphics[height=1.5in]{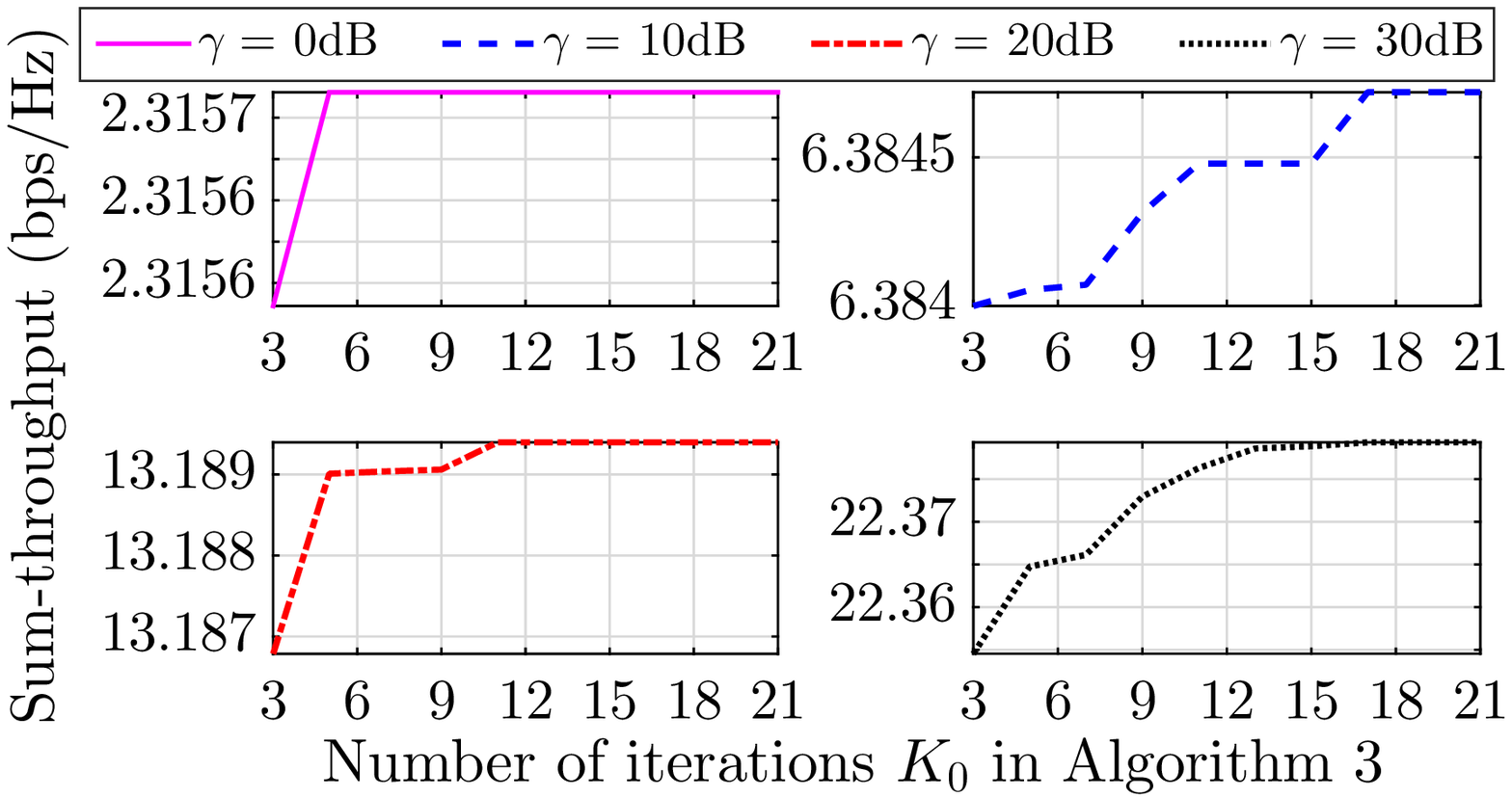}}}
	\caption{\small Validation of the proposed analysis and low-complexity claims regarding the joint optimization.}\label{fig:valid} 
\end{figure*}   
Furthermore, it is worth noting that since the EH devices, in contrast to the tags, have their own RF chain, TRX, and radio  signal generation unit, they can involve more sophisticated signal designing, instead of just controlling a scalar BC resembling power control. In fact as shown in~\cite{SRM-TVT-WPCN} in multiantenna EH users setting, one needs to design the precoder for them. Also, in WPCN, the optimal time for energy and information transfer phases needs to be optimized.

\begin{remark}	
	\textit{We can summarize that the proposed TRX designs for $\mathcal{R}$ (both individually-optimal and asymptotically-joint-optimal ones) hold equally good for HAP to maximize sum-rate in multiantenna HAP-powered uplink transmission from multiple single antenna RF-EH users. }
\end{remark}

\section{Numerical Results}\label{sec:res} 
Here, we numerically evaluate the performance of our proposed TRX and BC designs. Unless explicitly stated,  we have used $N=4, M=4$, $P_T=30$dBm, $\sigma_{\mathrm{w}_{\mathcal{T}}}^2=\sigma_{\mathrm{w}_{\mathcal{R}}}^2=-170$dBm, ${\eta=0},K=10\,NM, K_0=15,\xi=10^{-6}$ and $\beta_i=\varpi d_i^{-\varrho},\forall i$, where $\varpi=\left(\frac{3\times 10^8}{4\pi f}\right)^2$ being the average channel attenuation at unit reference distance with $f=915$MHz \cite{TSP19} being TX frequency, $d_i$ is $\mathcal{R}$ to $\mathcal{T}_i$ distance, and $\varrho=3$ is the path loss exponent. Noting the practical settings for the BC coefficients~\cite{QAM-BSC} as $\max\left\lbrace\abs{\mathrm{x}_{\mathcal{T}_k}}\right\rbrace=0.78$~\cite{BSC-Cascaded} and $\mathbb{E}\left\lbrace\abs{\mathrm{x}_{\mathcal{T}_k}}\right\rbrace=0.3162$~\cite{Bistatic-BCS}, we set $\alpha_{\min}=0.1\left(\mathbb{E}\left\lbrace\abs{\mathrm{x}_{\mathcal{T}_k}}\right\rbrace\right)^2=0.01$ and $\alpha_{\max}=\max\left\lbrace\abs{\mathrm{x}_{\mathcal{T}_k}}\right\rbrace\left(\mathbb{E}\left\lbrace\abs{\mathrm{x}_{\mathcal{T}_k}}\right\rbrace\right)^2=0.078,\forall k\in\mathcal{M}$. Regarding deployment, $M$ tags have been placed uniformly over a square field with length $L=100$m and $\mathcal{R}$ is placed at its center. While investigating individual optimizations, we have used fixed TX precoding as $\mathbf{f}_{\rm L}$ (EB design),  {combiner} as $\mathbf{G}_{\rm H}$ (ZF-based RX beamforming design), and BC vector as $\boldsymbol{\alpha}_{\rm H}$ (full-reflection mode).  Lastly, all the sum-backscattered-throughput performance results have been obtained numerically after averaging over $10^3$ independent channel realizations.

\begin{figure*}[!t]
\centering  
\subfigure[Optimal TX precoding.]{{\includegraphics[height=1.5in]{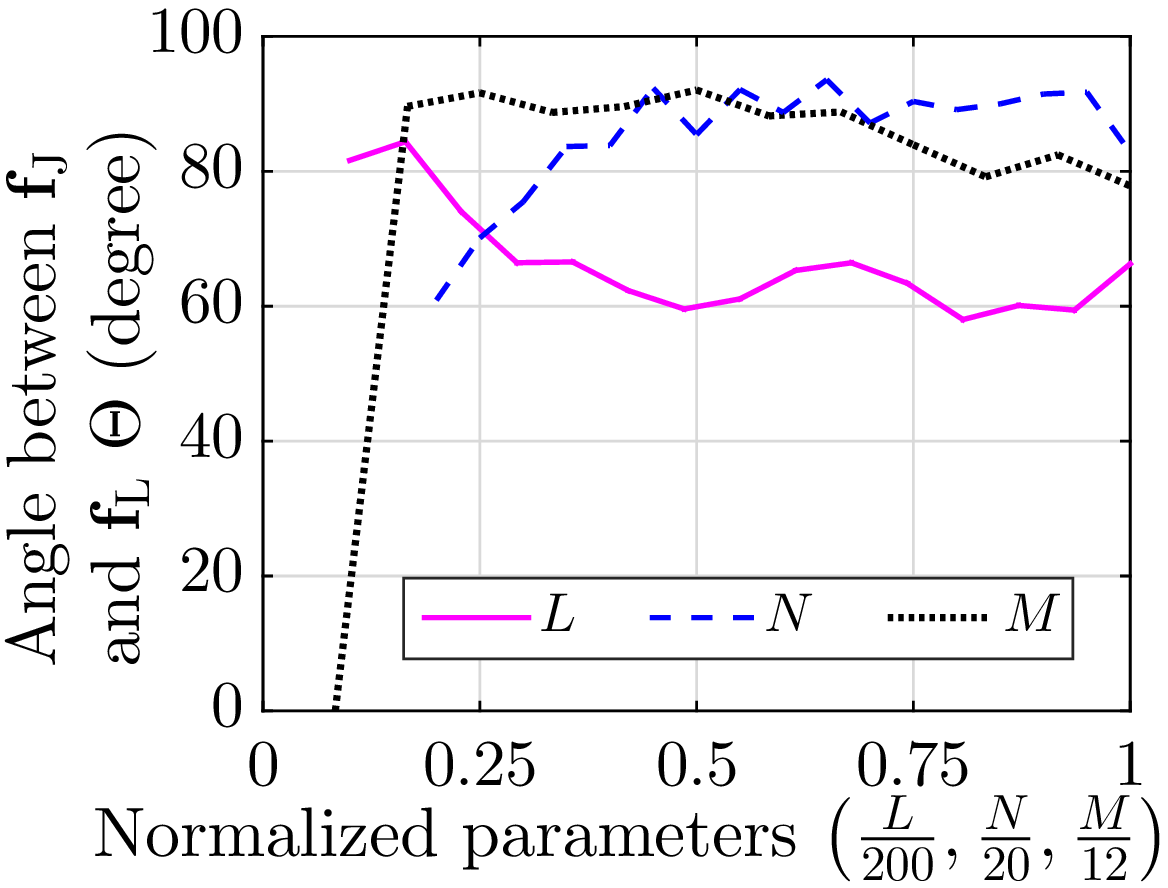} }}\quad
\subfigure[Asymptotically-optimal solutions.]{{\includegraphics[height=1.5in]{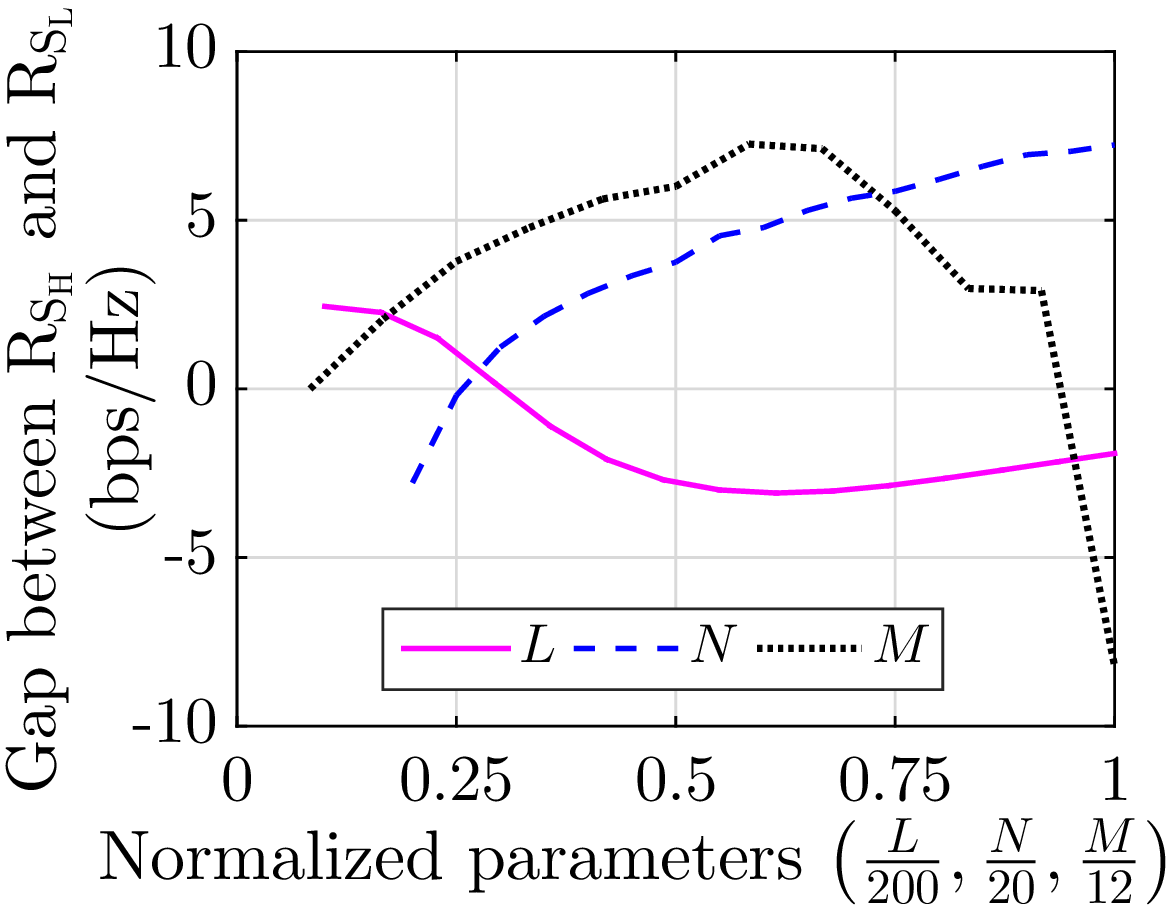}}}\quad 
\subfigure[Optimal BC designs.]{{\includegraphics[height=1.5in]{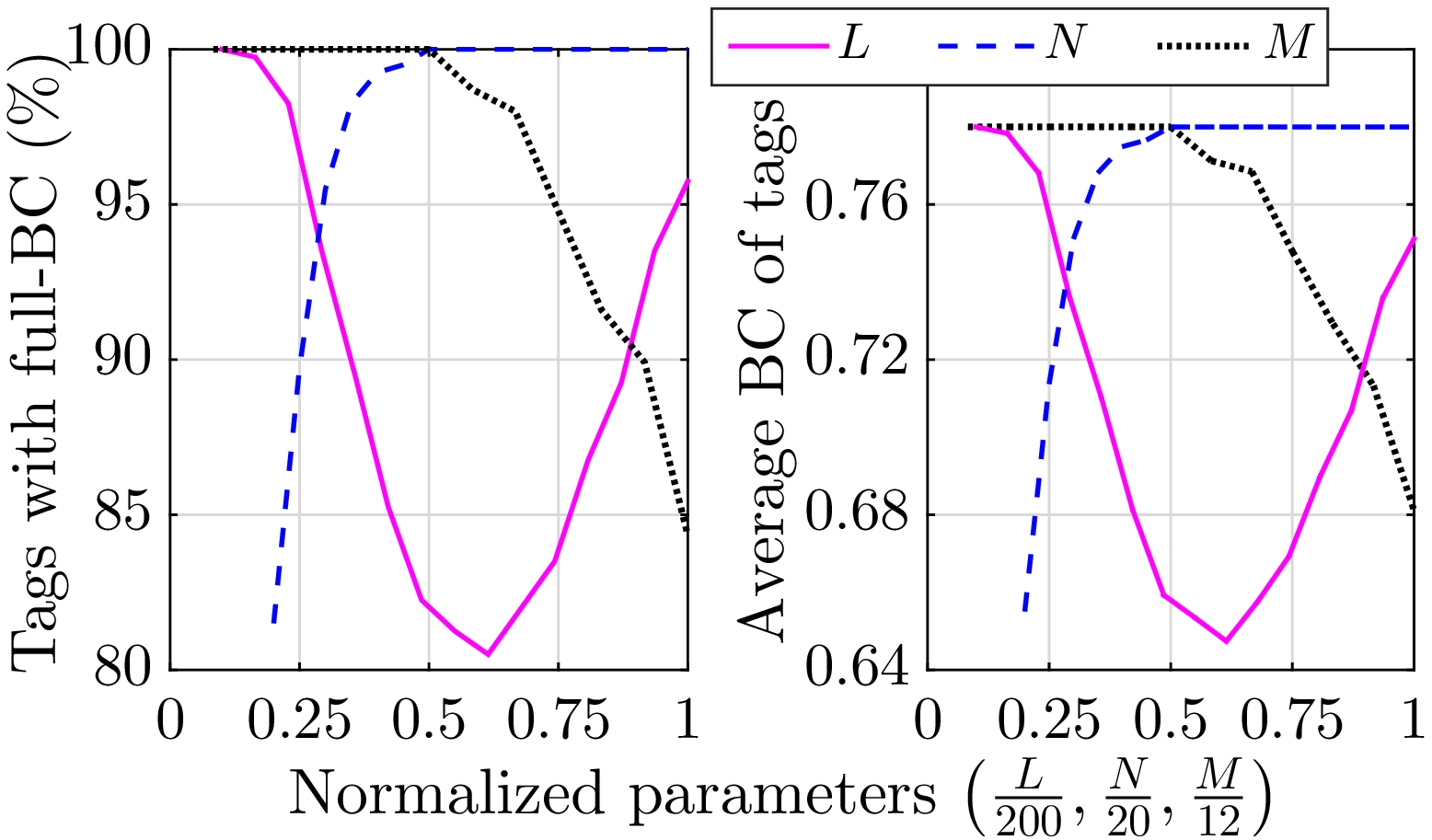} }} 
\caption{\small Key numerical insights on the proposed optimal TRX and BC designs.}\label{fig:insights} 
\end{figure*}
\subsection{Verification of Low-Complexity Designs} 
First we verify the quality of the proposed low-complexity designs against the joint TRX-BC design $\left(\mathbf{f}_{\rm J},\mathbf{G}_{\rm J},\boldsymbol{\alpha}_{\rm J}\right)$ as returned by Algorithm~\ref{Algo:Opt} for different effective backscattered SNR $\gamma\triangleq\frac{P_T\overline{\beta}^2}{\sigma_{\mathrm{w}_{\mathcal{R}}}^2}$ values, where $\overline{\beta}^2=\frac{1}{M}\sum_{i\in\mathcal{M}}\beta_i^2$. Specifically, the three low-complexity designs investigated in Fig.~\ref{fig:valid}(a) are: (i) \textit{optimal TRX design} $\left(\mathbf{f}_{\rm J{\alpha_H}},\mathbf{G}_{\rm J{\alpha_H}}\right)$ for fixed BC vector $\boldsymbol{\alpha}=\boldsymbol{\alpha}_{\rm H}$ as defined by Remark~\ref{rem:TRX}, (ii) asymptotically optimal joint TRX-BC design $\left(\mathbf{f}_{\mathrm H},\mathbf{G}_{\rm H},\boldsymbol{\alpha}_{\rm H}\right)$ for high-SNR applications as defined by Remark~\ref{rem:high}, and (iii) joint design $\left(\mathbf{f}_{\rm L},\mathbf{G}_{\rm L}\triangleq\left[\mathbf{g}_{\rm L_1}\;\mathbf{g}_{\rm L_2}\;\ldots\;\mathbf{g}_{\rm L_M}\right],\boldsymbol{\alpha}_{\rm L}\right)$ for low-SNR applications as summarized by Remark~\ref{rem:low}. It can be easily verified that all three  low-complexity designs closely follow the performance of Algorithm~\ref{Algo:Opt}. The exact sum-throughput gap between the one  achieved using Algorithm~\ref{Algo:Opt} and the ones with three low-complexity designs is also quantified in Fig.~\ref{fig:valid}(a). It is observed that the optimal TRX with $\boldsymbol{\alpha}=\boldsymbol{\alpha}_{\rm H}$ performs the best among three low-complexity designs with an average performance gap of $<0.04$ bps/Hz. Whereas, the low-SNR based joint design performs better for $\gamma\le34$dB. 

Next we focus on validating the fast convergence claim of Algorithm~\ref{Algo:Opt} by plotting the variation of the returned sum-backscattered-throughput $\mathrm{R}_{\mathrm{S}_{\rm J}}$ as against the increasing number of iterations $K_0$ in Fig.~\ref{fig:valid}(b) for different backscattered SNR values $\gamma$. {Here, we would like to highlight that the first result in Fig.~\ref{fig:valid}(b) is plotted for $K_0=3$ because with that we were able to consider the sum-backscattered-throughput of the best NM returned solution among the three starting precoder values, namely $\mathbf{f}_{\mathrm{L}},\mathbf{f}_{\mathrm{H}},$ and $\frac{ \mathbf{f}_{\mathrm{L}}+ \mathbf{f}_{\mathrm{H}}}{\norm{ \mathbf{f}_{\mathrm{L}}+ \mathbf{f}_{\mathrm{H}}}}.$ Since, these three points cover the entire feasible range (left, right, and center) for the weighted precoding vector ${\mathbf{f}}_{\rm w_0}$ defined by 
\eqref{eq:wTXB0}, the relative gap between the achievable throughput for $K_0=3$ and other higher iterations is not very significant. This gap between sum-throughputs for $K_0=3$ and higher iterations, say $K_0=15$, is practically meaningful only for high SNR scenarios, like $\gamma=30$dB.}  Also, it can be verified that $K_0=5$ iterations are in general sufficient for achieving an acceptable performance quality versus computational complexity tradeoff and the throughput enhancement beyond $K_0\ge15$ is negligible.

\begin{figure*}[!t]
\centering  
\subfigure[Variation with field size.]{{\includegraphics[height=1.4in]{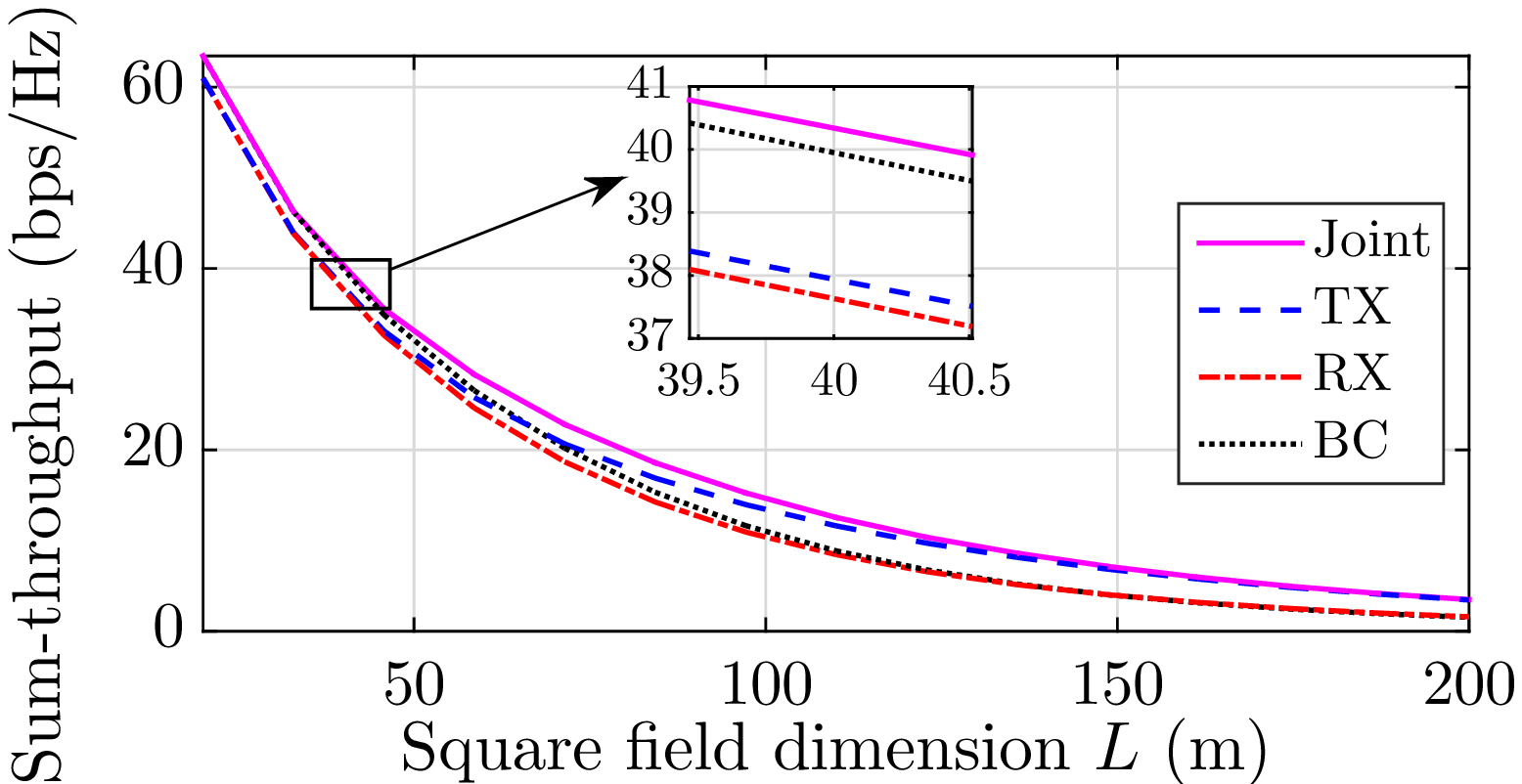} }}\quad 
\subfigure[Variation with array size.]{{\includegraphics[height=1.4in]{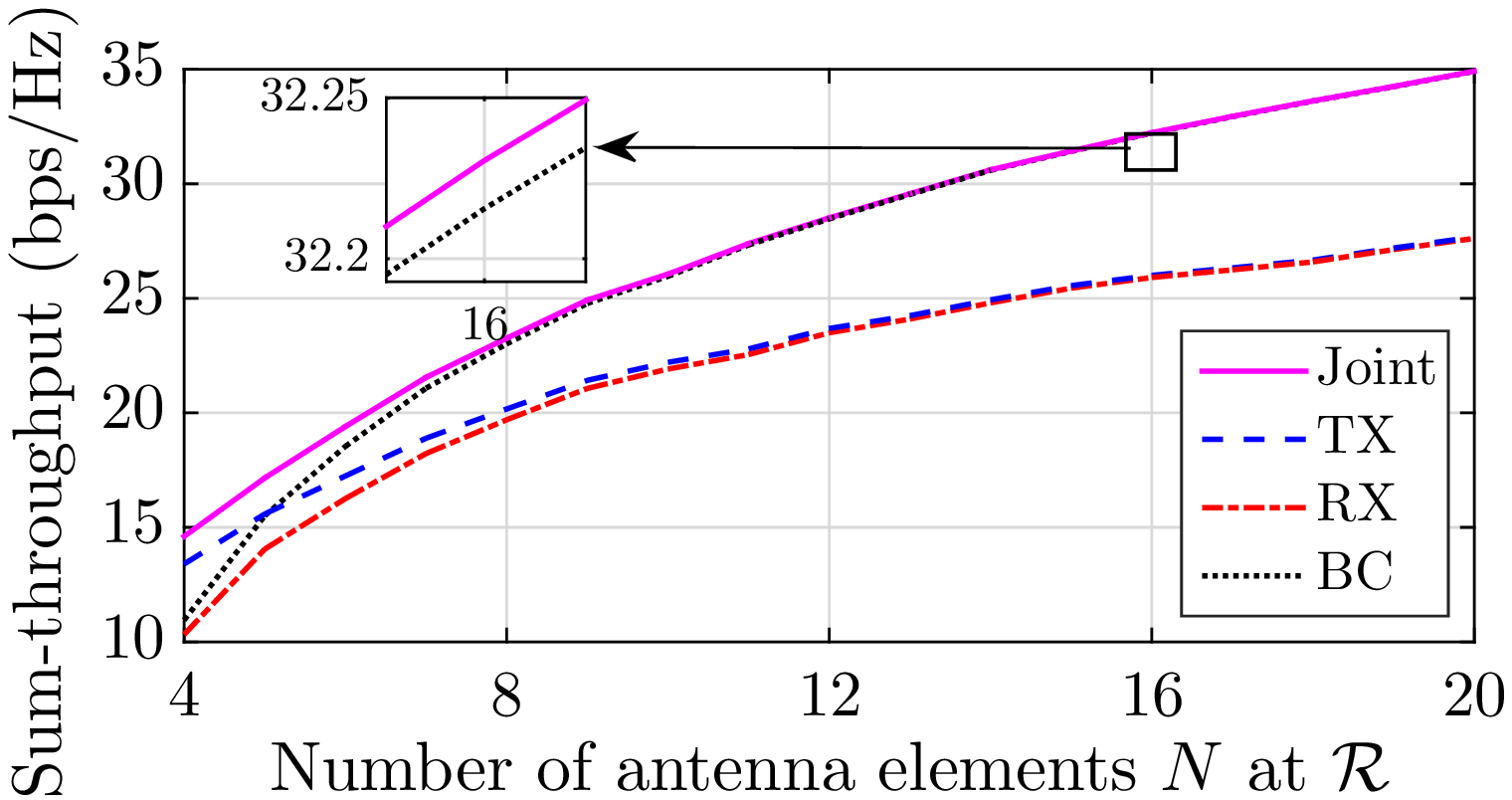}}} 
\subfigure[Variation with population of tags.]{{\includegraphics[height=1.4in]{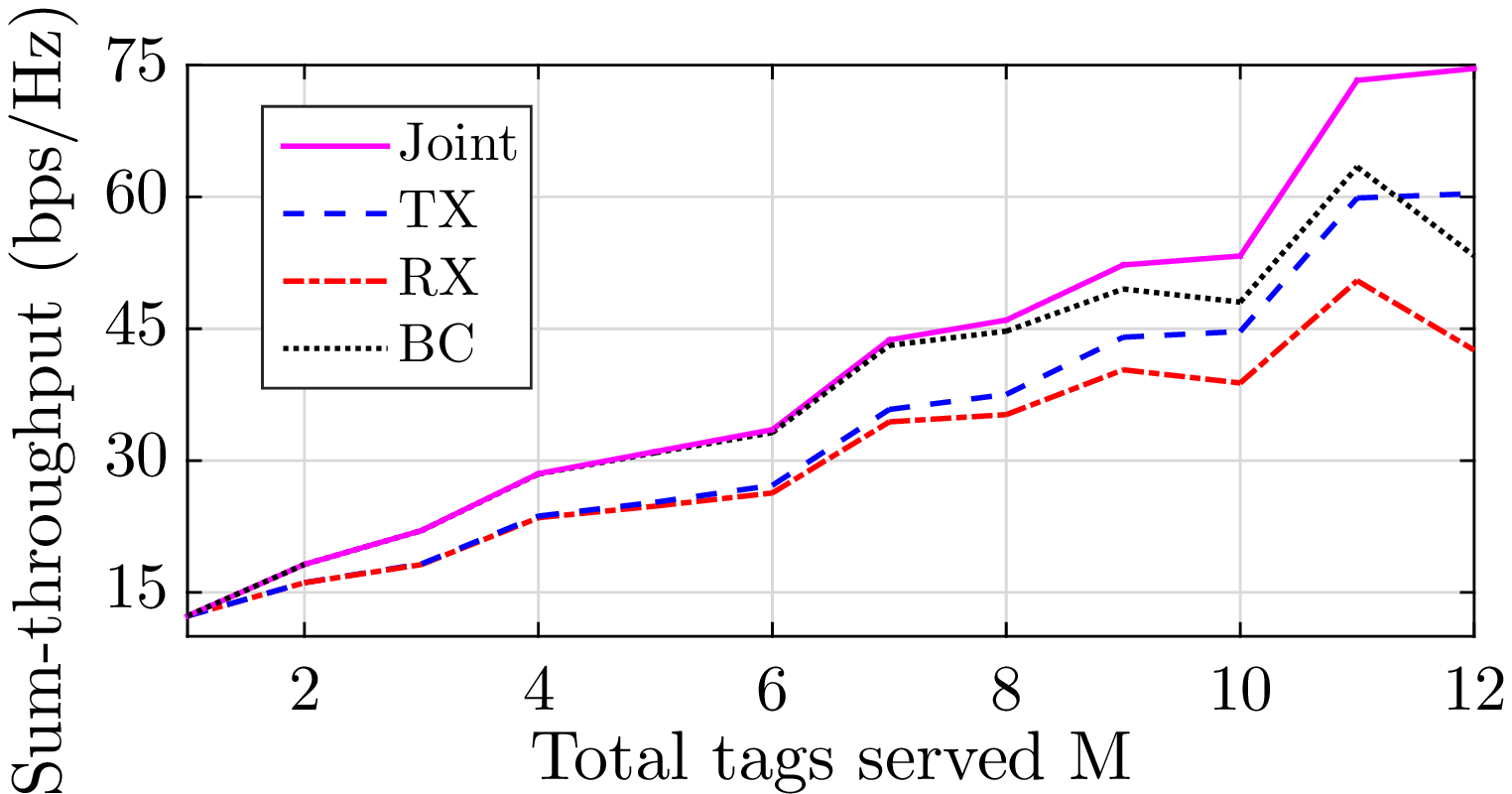}}}\quad 
\subfigure[{Variation with CSI imperfection parameter $\eta$.}]{{\includegraphics[height=1.4in]{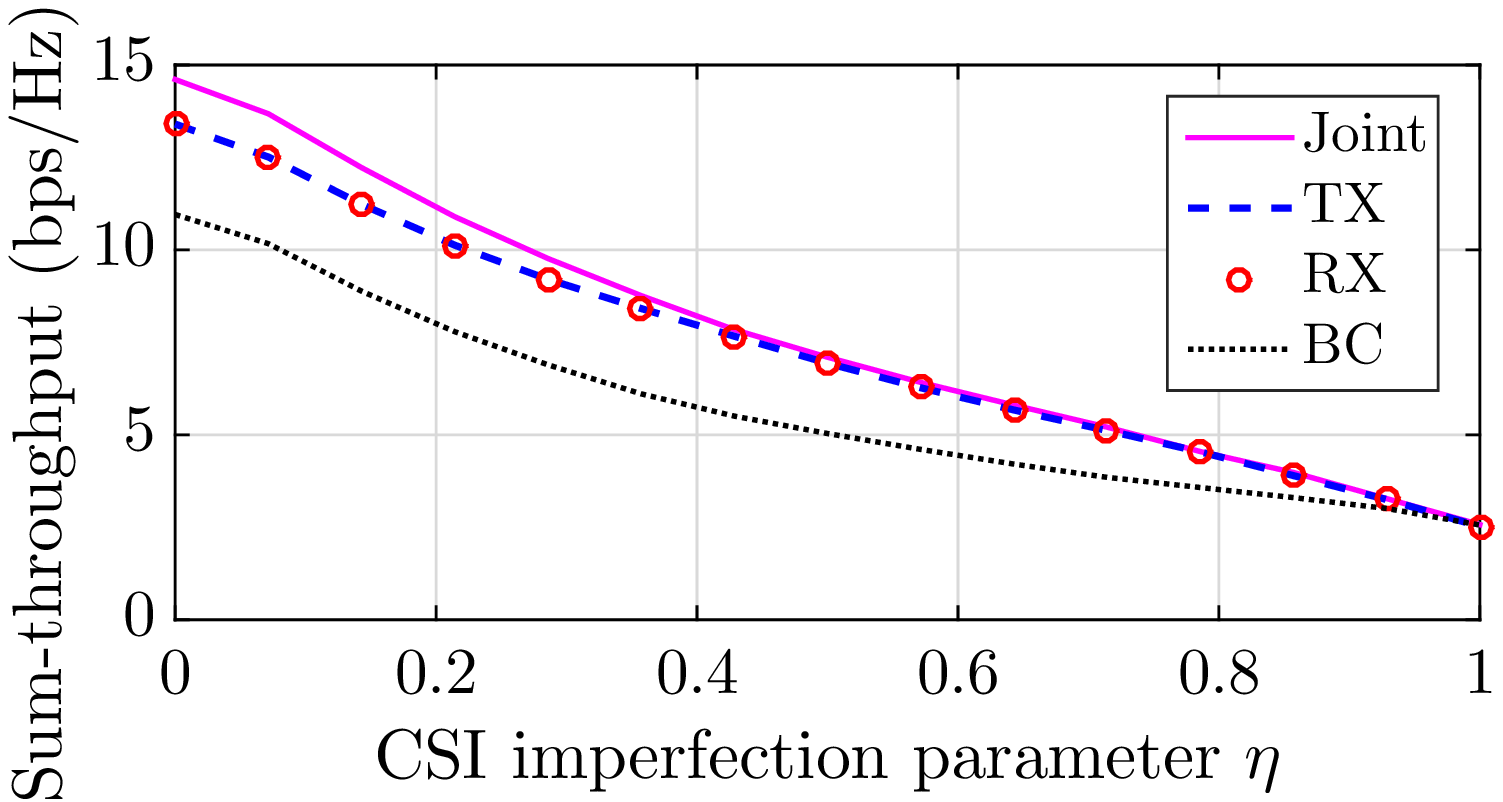} }}   
\caption{\small {Investigating the relative performance of the proposed individually-optimal designs against the jointly-optimal one.}}\label{fig:indiv} 
\end{figure*}  
\subsection{Key Insights on Optimal TRX-BC Design}
Here we present key features of the proposed optimal designs for varying different system parameter values as: (a) field size $L$ between $20$m and $200$m, (b) number $N$ of antennas at $\mathcal{R}$ between $4$ and $20$, and (c) number $M$ of tags between $1$ and $12$. We start with first plotting the angle between the proposed optimal precoder $\mathbf{f}_{\rm J}$, as returned by Algorithm~\ref{Algo:Opt}, and the TX-EB vector $\mathbf{f}_{\rm L}$, which is asymptotically-optimal under low-SNR regime, in Fig.~\ref{fig:insights}(a). From this result we observe that the angle $\Theta\triangleq\cos^{-1}\left\lbrace\frac{\mathrm{real}\left\lbrace\mathbf{f}_{\rm J}^{\rm H}\,\mathbf{f}_{\rm L}\right\rbrace}{\norm{\mathbf{f}_{\rm J}}\,\norm{\mathbf{f}_{\rm L}}}\right\rbrace$ between the directions $\mathbf{f}_{\rm J}$ and $\mathbf{f}_{\rm L}$ remains similar through the respective variation of the three system parameters $L,N,$ and $M$. This result shows that the TX beamforming direction $\mathbf{f}_{\rm J}$ maximizing the sum-backscattered throughput from the tags is quite different from  $\mathbf{f}_{\rm L}$ that maximizes the sum-received power among the tags.
 
Next via Fig.~\ref{fig:insights}(b), we investigate the relative superiority of the two asymptotically-optimal designs. Apart for  $M=1$, where since both the asymptotically-optimal designs are identical (i.e., the TRX and BC design being respectively in MRT-MRC and full-refection modes), the throughput performances are same, for the other parameter values the high-SNR design becomes superior to the low-SNR one after a certain SNR threshold. Specifically, the sum-backscattered-throughput $\mathrm{R_{S_H}}\triangleq\mathrm{R_S^o}\left(\mathbf{f}_{\mathrm{H}},\mathbf{G}_{\mathrm{H}},\boldsymbol{\alpha}_{\mathrm{H}}\right)$ for the high-SNR based design is larger than its corresponding counterpart
$\mathrm{R_{S_L}}\triangleq\mathrm{R_S^o}\left(\mathbf{f}_{\mathrm{L}},\mathbf{G}_{\mathrm{L}},\boldsymbol{\alpha}_{\mathrm{L}}\right)$ for $L\le70$m, $N\ge6$, and $M\le11$.  
\begin{remark}\label{rem:insights}
	\textit{For large antenna array at $\mathcal{R}$, the joint design is characterized by the ZF-based {combiner} $\mathbf{G}_{\rm H}$ and full-reflection mode $\boldsymbol{\alpha}_{\mathrm{H}}$ based BC  design with optimal precoder $\mathbf{f}_{\rm H}$ maintaining the balance between individual MRT direction for each tag. Whereas, the EB based TX precoding $\mathbf{f}_{\rm L}$, MMSE-based {combiner} $\mathbf{G}_{\rm L}$, and binary BC design $\boldsymbol{\alpha}_{\mathrm{L}}$, which is asymptotically optimal under low-SNR regime, can be preferred for larger BSC field-sizes and denser tag-deployments.}
\end{remark}  
{Also, remember that relatively  higher values of BSC range~\cite{New-SP-Mag2} have been used in the simulations due to the consideration of  multiantenna reader~\cite{TSP19,Impinj} and semi-passive tags~\cite{semi-passive}.}

Lastly, we shed novel key insights on the optimal BC design by plotting the variation of: (i) total fraction of tags (in $\%$) following the full-reflection mode (i.e., $\alpha_k=\alpha_{\rm max}$), and (ii) average BC value $\frac{1}{M\left(\mathbb{E}\left\lbrace\abs{\mathrm{x}_{\mathcal{T}_k}}\right\rbrace\right)^2}\sum\limits_{i\in\mathcal{M}}\alpha_i=\frac{10}{M}\sum\limits_{i\in\mathcal{M}}\alpha_i$ among the tags, for different values of $L,N,$ and $M$ in Fig.~\ref{fig:insights}(c). We can observe that for the BSC applications having larger antenna array ($N\ge10$) at $\mathcal{R}$ to serve relatively smaller number of tags $M\le6$, almost all the tags (as represented by $100\%$) set their BC in the full-refection mode (plotted as $0.78$). Furthermore, in general,  $>80\%$ tags prefer the full-reflection mode as their BC design. The average value of the BC design among the tags also follow a very similar trend, corroborating the fact that optimal BC designing involves  a \textit{low-complexity binary decision-making process} (cf. Remark~\ref{rem:low}).

\begin{figure*}[!t]
	\centering  
	\subfigure[Variation with field size.]{{\includegraphics[height=1.4in]{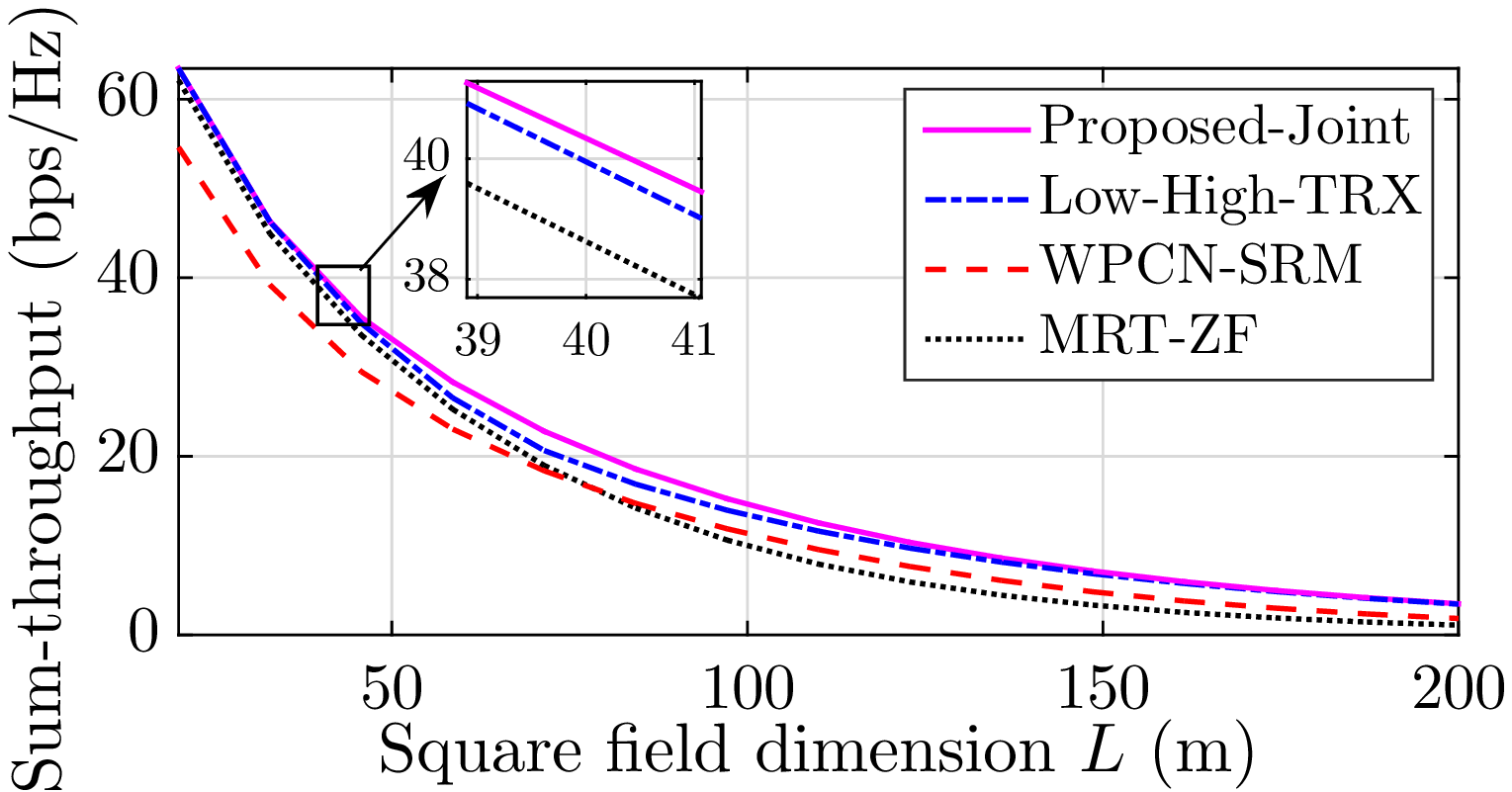} }}\quad
	\subfigure[Variation with array size.]{{\includegraphics[height=1.4in]{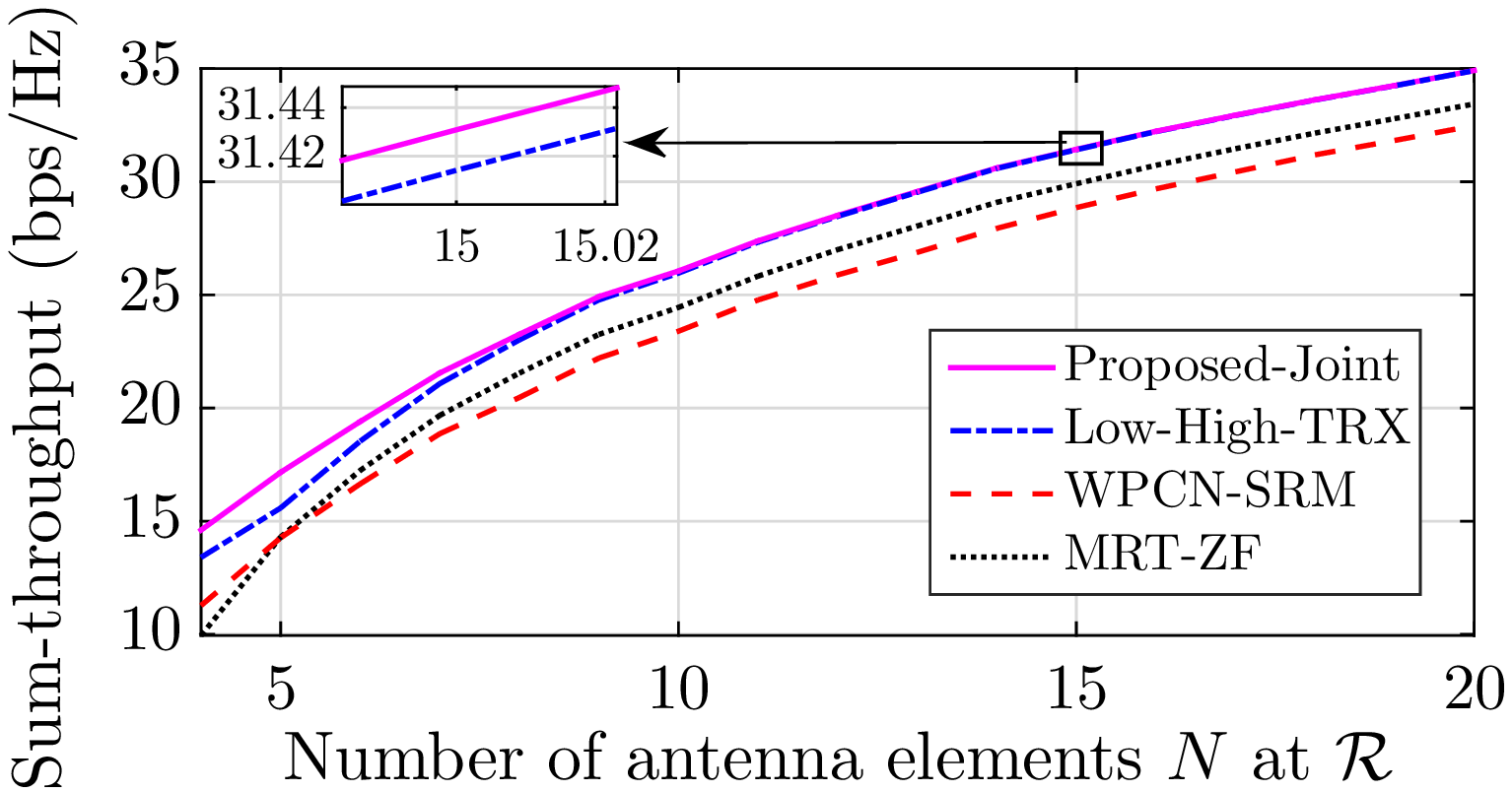}}} 
	\subfigure[Variation with population of tags.]{{\includegraphics[height=1.4in]{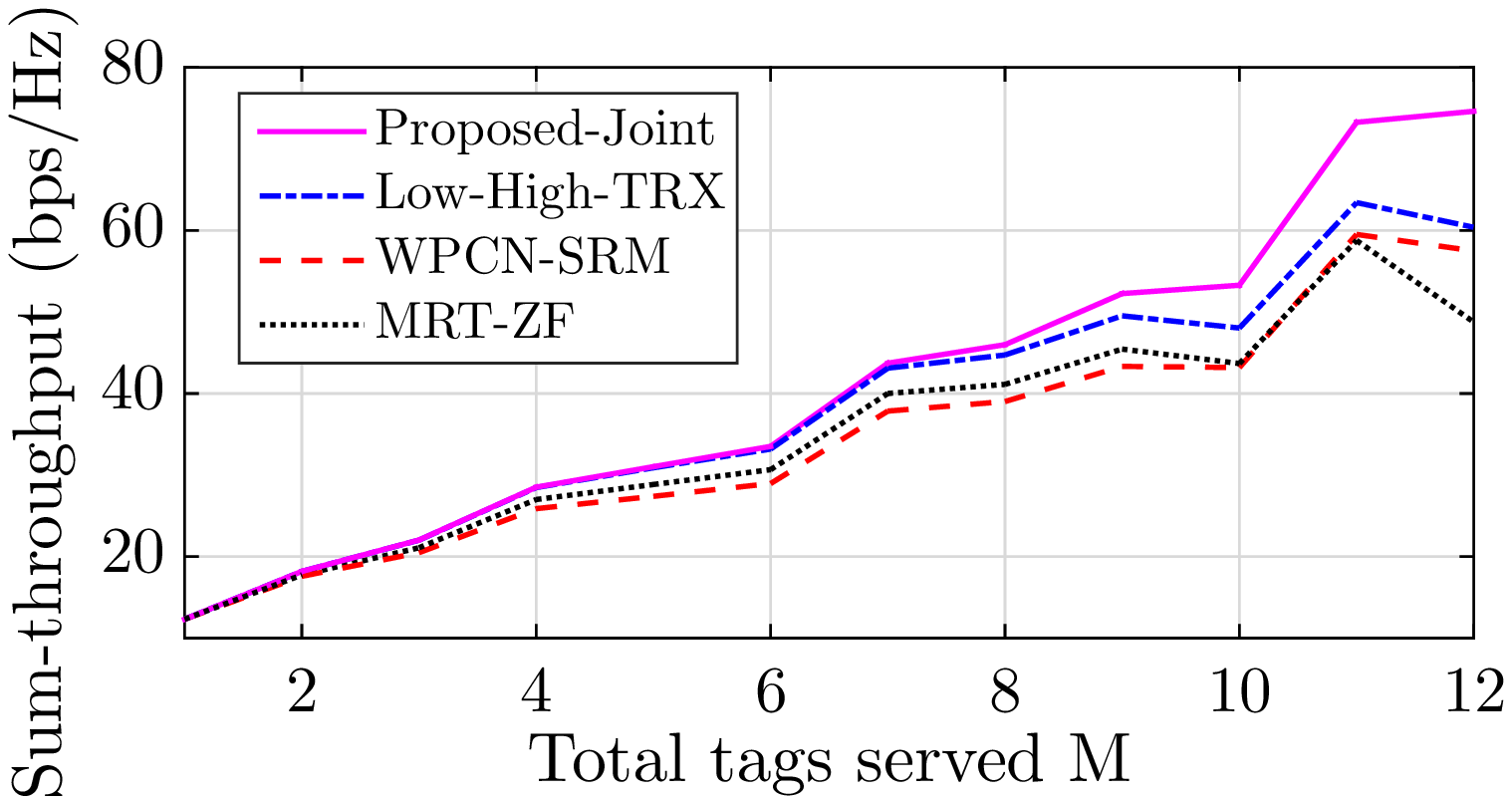} }}\quad 
	\subfigure[{Variation with CSI imperfection parameter $\eta$.}]{{\includegraphics[height=1.4in]{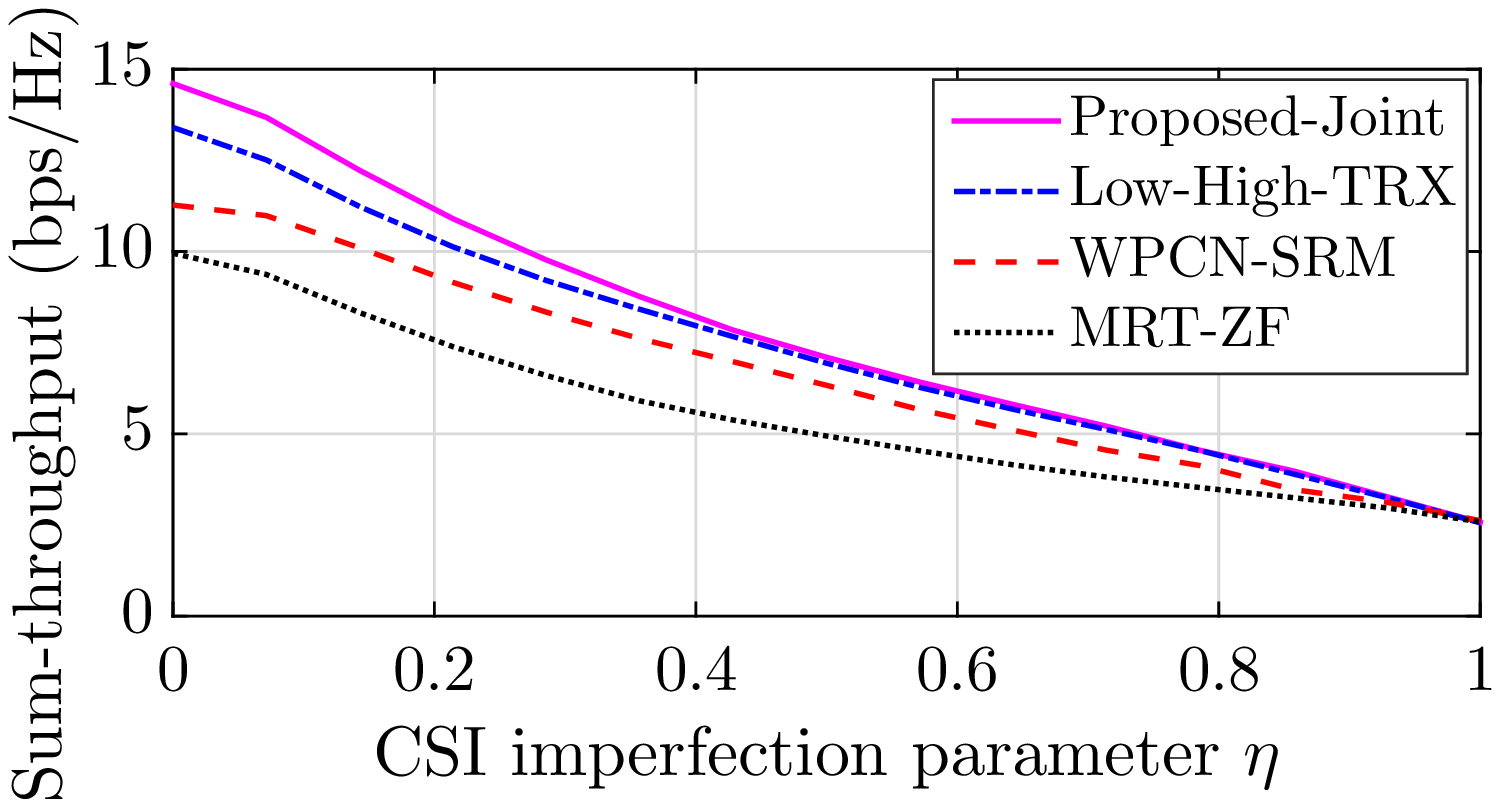} }}   
	\caption{\small {Sum-throughput performance comparison of the proposed joint TRX and BC design against the relevant benchmarks~\cite{WPCN-MIMO-SRM,GIoT-BSC}.}}\label{fig:comp} 
\end{figure*}       
\subsection{Comparison Among Semi-Adaptive Schemes}\label{sec:Comp-Semi} 
In this section we conduct a relative performance comparison study among the three semi-adaptive designs and the fully-adaptive one as obtained using Algorithm~\ref{Algo:Opt}. The three semi-adaptive designs involving individual optimizations of TX precoding, RX beamforming, and BC vector, respectively, are plotted in Figs.~\ref{fig:indiv}(a),~\ref{fig:indiv}(b), and~\ref{fig:indiv}(c)  along with the joint design for varying $L,N,$ and $M$. From Fig.~\ref{fig:indiv}(a), we notice that the optimal TX precoding with fixed $\mathbf{G}=\mathbf{G}_{\rm H}$ and  $\boldsymbol{\alpha}=\boldsymbol{\alpha}_{\rm H}$ performs better than the other two semi-adaptive designs for higher values of $L$  with $N=M=4$ and eventually approaches the throughput performance of joint design. Whereas, with optimal RX beamforming design being the weakest scheme as observed  in Fig.~\ref{fig:indiv}, implies that MMSE-based design is not that critical and in fact ZF-based asymptotically-optimal one is practically good enough. Furthermore, the optimal BC design having TX-EB as precoder and ZF-based RX-beamforming turns out to be the best semi-adaptive scheme except under very low SNR regimes as represented via $L\ge70$m, $N\le5$, and $M\ge12$ in Figs.~\ref{fig:indiv}(a),~\ref{fig:indiv}(b), and~\ref{fig:indiv}(c), respectively. Thus, TX precoding optimization is relatively more critical than RX beamforming, and closely follows the optimal BC design. Overall, the average improvement of the joint design over TX-precoding, RX-beamforming, and BC designs is respectively $12.5\%$, $23.5\%$, and $11.5\%$.

Other than the above comparisons for perfect CSI case, we have also investigated the relative robustness of proposed designs against inaccuracy in the available CSI via Fig.~\ref{fig:indiv}(d). Here it is observed that with the CSI imperfection parameter $\eta$ increased from $0$ (perfect CSI case) to $1$ (statistical CSI case), the average sum-backscattered-throughput for the joint, TX (or RX), and BC optimization schemes respectively decreases by $7.6$dB, $7.3$dB, and $6.4$dB. Despite this performance loss due to the inaccuracy of the available CSI, the proposed jointly-optimal TRX-BC  design provides an average  enhance of about $3.9\%$ over the optimal TX precoding and RX beamforming designs, which have almost the same performance. Whereas, this performance gain over optimal BC is around $23\%$. Another important note from Fig.~\ref{fig:indiv}(d) is that the relatively high gap between the jointly and individually (TX, RX, BC) optimal designs at $\eta=0$, diminishes with increasing $\eta$ and for $\eta=1$ case  implying no instantaneous CSI information, all the schemes have almost the same performance. This corroborates the need for having an accurate CSI at $\mathcal{R}$.

\subsection{Achievable Throughput Gains over Benchmarks}\label{sec:bench}
Finally, to corroborate the practical utility of the proposed designs, we here conduct a performance comparison study against the two available benchmark designs, namely, (i) WPCN-SRM scheme~\cite{WPCN-MIMO-SRM}  targeted towards the TRX designing at the multiantenna HAP for the uplink sum-rate-maximization (SRM) from the multiple single-antenna EH users, and (ii) MRT-ZF scheme~\cite{GIoT-BSC,BSC-WET-MIMO} where the MRT based precoder and  ZF based {combiner} are designed for each tag. As both these benchmarks do not consider BC optimization and use $\boldsymbol{\alpha}=\boldsymbol{\alpha}_{\rm H}$, apart from our proposed joint design obtained using Algorithm~\ref{Algo:Opt}, we have also included our proposed asymptotically-optimal scheme (denoted by Low-High-TRX in Fig.~\ref{fig:comp}) where we have chosen the TRX design with $\boldsymbol{\alpha}=\boldsymbol{\alpha}_{\rm H}$ between $\left(\mathbf{f}_{\rm L},\mathbf{G}_{\rm L}\right)$ and $\left(\mathbf{f}_{\rm H},\mathbf{G}_{\rm H}\right)$ based on whether $\mathrm{R_{S_L}}\geq\mathrm{R_{S_H}}$, or not.

The performance comparison results for the two proposals against the two benchmarks are plotted in Figs.~\ref{fig:comp}(a),~\ref{fig:comp}(b), and~\ref{fig:comp}(c)  for varying $L,N,$ and $M$, respectively, assuming perfect CSI ($\eta=0$) at $\mathcal{R}$. {It is clearly visible that both  jointly-optimal (cf. Algorithm~\ref{Algo:Opt}) and   asymptotically-optimal designs \textit{outperform} both the benchmarks. The low-high-SNR based TRX design with BC $\boldsymbol{\alpha}_{\rm H}$ yielding sum-backscattered-throughput as $\max\left\lbrace\mathrm{R_{S_L}},\mathrm{R_{S_H}}\right\rbrace$,  respectively provides an \textit{average improvement} of about $18\%$ and $28\%$ over the  WPCN-SRM and MRT-ZF schemes in terms of achievable sum-backscattered-throughput. The main reasons for this  significant improvement are that the TX-EB based common precoding design performs much better in terms of sum-throughput  than the respective MRT design for each tag as proposed in~\cite{GIoT-BSC,BSC-WET-MIMO}, and than  TX precoder of WPCN-SRM scheme which is aimed at optimizing a non-equivalent goal defined in~\cite[Prop. 1]{WPCN-MIMO-SRM}. Furthermore, the proposed TRX-BC design $\left(\mathbf{f}_{\rm J},\mathbf{G}_{\rm J},\boldsymbol{\alpha}_{\rm J}\right)$ provides an \textit{additional throughput enhancement} of $4\%$ over the low-high-SNR-based TRX design.}

{Lastly, to corroborate the practical utility of the proposed joint TRX-BC designs under non-availability perfect CSI, we have also conducted this comparison for different $\eta$ in Fig.~\ref{fig:comp}(d). On an average, over all the possible values of $\eta$,  the proposed joint-TRX-BC design provides an \textit{average improvement} of about $4\%, 15\%,$ and $37\%$ over the Low-High-TRX, WPCN-SRM and MRT-ZF schemes, respectively. However, this performance enhancement of the proposed joint TRX-BC scheme diminishes with increasing $\eta$ denoting  inaccuracy in the available CSI. In fact these respective  gains decrease from $9\%, 29\%,$ and $46\%$ for $\eta=0$ (i.e., accurate perfectly CSI) to $3\%, 12\%,$ and $43\%$ for $\eta=0.5$, before finally reducing to zero for $\eta=1$. Thus, the full potential of the proposed schemes is realized when a relatively accurate CSI is available at $\mathcal{R}$.} 
   
\section{Concluding Remarks}\label{sec:concl}   
This work investigated novel sum-backscattered-throughput maximization problem that jointly optimizes the TRX design at the multiantenna reader and BC at the single antenna tags. Noting the non-convexity  of joint-optimization problem, novel generalized-convexity principles~\cite{avriel2010generalized} were explored to obtain  the individually optimal TX and RX beamforming designs while reducing the BC optimization to a power control problem  over the multiple interfering links. Further, while exploring the asymptotically-optimal joint designs in both low and high SNR regimes, we discoursed that the optimal TX precoding vector is based on the direction that trade-offs between the one maximizing the sum-received RF power among the tags and one balancing among the individual MRT direction for each tag. Whereas the {combiner} design is based on MMSE beamforming and the BC optimization reduces to a low-complexity binary decision-making process. Detailed numerical investigation validating the fast convergence claims of the proposed iterative algorithm for the joint design and near-optimal performance of the asymptotically-optimal low-complexity designs, showed that the proposed solutions can yield an overall $20\%$ enhancement over the  benchmarks. Lastly from the system-engineering perspective, this work providing key insights on the optimal TRX-BC design for multi-tag monostatic MIMO-BSC settings, also discloses how these solutions can be utilized for other  applications like WPCN and ambient or bi-static BSC settings with multiantenna tags.

\appendices
\setcounter{equation}{0}
\setcounter{figure}{0}
\renewcommand{\theequation}{A.\arabic{equation}}
\renewcommand{\thefigure}{A.\arabic{figure}}
\section{Proof of Lemma~\ref{lem:prec}: Property of Optimal Precoders}\label{app:prec}  
From $\mathcal{O}_{\mathrm{S}}$, we note that for a given {combiner} $\mathbf{g}_k$ and BC  design $\alpha_k$ for each $\mathcal{T}_k$, the optimal precoders  $\mathcal{T}_k$ are characterized by the ones maximizing the Lagrange function $\mathcal{L}_{\mathrm{T}}$ with respect to $\mathbf{f}_k$, where $\mathcal{L}_{\mathrm{T}}$ is defined below and $\nu_{\mathrm{T}}\ge0$ is Lagrange multiplier for $({\rm C1})$:	
\begin{eqnarray}
\mathcal{L}_{\mathrm{T}}=\sum_{k\in\mathcal{M}} \log_2\left(1+\gamma_{\mathcal{R}_k}\right)+ \nu_{\mathrm{T}}\left(P_T- \sum_{k\in\mathcal{M}}\lVert\mathbf{f}_k\rVert^2\right).
\end{eqnarray}

{As a result,} the optimal precoders can be obtained by solving their below respective subgradient Karush Kuhn Tucker (KKT) condition~\cite[Ch. 5.5.3]{boyd} in terms of $\mathbf{f}_k$ for each $\mathcal{T}_k$:	
\begin{align} \label{eq:subgT}
\frac{\partial\mathcal{L}}{\partial\mathbf{f}_k}
=\sum\limits_{m\in\mathcal{M}}\mathbf{Z}_m\,\mathbf{f}_{k}-\nu_{\mathrm{T}}\,\mathbf{f}_{k}=\mathbf{0}_{N\times1}.
\end{align}
where $\mathbf{Z}_m\triangleq\frac{\alpha_m\left|\mathbf{g}_{m}^{\rm H}\mathbf{h}_{m}\right|^2\mathbf{h}_{m}^*\,\mathbf{h}_{m}^{\rm T}  -\gamma_{\mathcal{R}_m}\sum\limits_{i\in\mathcal{M}_m}\alpha_i\left|\mathbf{g}_{m}^{\rm H}\,\mathbf{h}_{i}\right|^2\,\mathbf{h}_{i}^*\,\mathbf{h}_{i}^{\rm T} }{\nu_{\mathrm{T}}\ln\left(2\right)\left( \sum\limits_{i\in\mathcal{M}}\alpha_i\,\left|\mathbf{g}_{m}^{\rm H}\,\mathbf{h}_{i}\right|^2\,\left|\mathbf{h}_{i}^{\rm T}\sum\limits_{{\overline{m}}\in\mathcal{M}}\mathbf{f}_{\overline{m}}\right|^2+\sigma_{\mathrm{w}_{\mathcal{R}}}^2\norm{\mathbf{g}_{m}}^2 \right)},$ $\forall m\in\mathcal{M}$.
Furthermore, since \eqref{eq:subgT} can be rewritten as  $\mathbf{f}_{k}=\sum_{m\in\mathcal{M}}\mathbf{Z}_m\,\mathbf{f}_{k}$ for each $\mathcal{T}_k$. it proves that the optimal precoder, denoted by $\mathbf{f}\in\mathbb{C}^{N\times1}$, is identical for all  tags, and we can write $\mathbf{f}_k=\frac{1}{\sqrt{M}}\mathbf{f}$ for each $\mathcal{T}_k,$ so that  ({\rm C1}) is satisfied.

\setcounter{equation}{0}
\setcounter{figure}{0}
\renewcommand{\theequation}{B.\arabic{equation}}
\renewcommand{\thefigure}{B.\arabic{figure}}
\section{Proof of Lemma~\ref{lem:ccF}: Concavity of $\overline{\mathrm{R}}_{\mathrm{S}}$ in  $\boldsymbol{\mathcal{F}}$}\label{app:ccF}
Below we redefine the backscattered-SINR in \eqref{eq:SINR} as  $\overline{\gamma}_{\mathcal{R}_k}$ for each $\mathcal{T}_k$ using matrix  $\boldsymbol{\mathcal{F}}$ definition: 
\begin{equation}\label{eq:SINR-F}
\overline{\gamma}_{\mathcal{R}_k}\triangleq\frac{\alpha_k\,\left|\mathbf{g}_{k}^{\rm H}\,\mathbf{h}_{k}\right|^2\,
	{\mathbf{h}_{k}^{\rm T}\boldsymbol{\mathcal{F}}\,\mathbf{h}_{k}^*}}{\sum_{i\in\mathcal{M}_k}\alpha_i\left|\mathbf{g}_{k}^{\rm H}\,\mathbf{h}_{i}\right|^2\,
	{\mathbf{h}_{i}^{\rm T}\boldsymbol{\mathcal{F}}\,\mathbf{h}_i^*}+\sigma_{\mathrm{w}_{\mathcal{R}}}^2\norm{\mathbf{g}_{k}}^2}.
\end{equation}

Following the systematic theory of deriving the second-order complex differential of the real scalar function $\overline{\gamma}_{\mathcal{R}_k}$ with respect to a complex matrix variable $\boldsymbol{\mathcal{F}}$~\cite{complex-matrixbook}, we need to investigate the corresponding combined or bigger Hessian matrix $\mathbb{H}\big(\overline{\gamma}_{\mathcal{R}_k}\big)$ of size $2N^2\times 2N^2$, containing the four $N^2\times N^2$ Hessian matrices $\boldsymbol{\mathcal{H}}_{\boldsymbol{\mathcal{F}}\boldsymbol{\mathcal{F}}^*}\big(\overline{\gamma}_{\mathcal{R}_k}\big),\boldsymbol{\mathcal{H}}_{\boldsymbol{\mathcal{F}}^*\boldsymbol{\mathcal{F}}^*}\big(\overline{\gamma}_{\mathcal{R}_k}\big),\boldsymbol{\mathcal{H}}_{\boldsymbol{\mathcal{F}}\boldsymbol{\mathcal{F}}}\big(\overline{\gamma}_{\mathcal{R}_k}\big),$ and $\boldsymbol{\mathcal{H}}_{\boldsymbol{\mathcal{F}}^*\boldsymbol{\mathcal{F}}}\big(\overline{\gamma}_{\mathcal{R}_k}\big)$ as its sub-elements, which is defined as~\cite{complex-matrixbook}:
\begin{align}
\mathbb{H}\big(\overline{\gamma}_{\mathcal{R}_k}\big)=&
\left[ \begin{array}{ccc}
\boldsymbol{\mathcal{H}}_{\boldsymbol{\mathcal{F}}\boldsymbol{\mathcal{F}}^*}\big(\overline{\gamma}_{\mathcal{R}_k}\big)
&\boldsymbol{\mathcal{H}}_{\boldsymbol{\mathcal{F}}^*\boldsymbol{\mathcal{F}}^*}\big(\overline{\gamma}_{\mathcal{R}_k}\big) \\
\boldsymbol{\mathcal{H}}_{\boldsymbol{\mathcal{F}}\boldsymbol{\mathcal{F}}}\big(\overline{\gamma}_{\mathcal{R}_k}\big)
&\boldsymbol{\mathcal{H}}_{\boldsymbol{\mathcal{F}}^*\boldsymbol{\mathcal{F}}}\big(\overline{\gamma}_{\mathcal{R}_k}\big)  \end{array} \right]\nonumber\\
=&\left[ \begin{array}{cc}
\frac{\partial}{\partial\boldsymbol{\mathcal{F}}}\left(\mathrm{vec}^{\rm T}\left\lbrace\mathrm{vec}\left\lbrace\frac{\partial\overline{\gamma}_{\mathcal{R}_k}}{\partial\boldsymbol{\mathcal{F}}^*}\right\rbrace\right\rbrace\right) & \\
\frac{\partial}{\partial\boldsymbol{\mathcal{F}}}\left(\mathrm{vec}^{\rm T}\left\lbrace\mathrm{vec}\left\lbrace\frac{\partial\overline{\gamma}_{\mathcal{R}_k}}{\partial\boldsymbol{\mathcal{F}}}\right\rbrace\right\rbrace\right) &  \end{array}\right.\nonumber\\
 &\qquad\quad\;\,\left. \begin{array}{cc}
 & \frac{\partial}{\partial\boldsymbol{\mathcal{F}}^*}\left(\mathrm{vec}^{\rm T}\left\lbrace\mathrm{vec}\left\lbrace\frac{\partial\overline{\gamma}_{\mathcal{R}_k}}{\partial\boldsymbol{\mathcal{F}}^*}\right\rbrace\right\rbrace\right)\\
 & \frac{\partial}{\partial\boldsymbol{\mathcal{F}}^*}\left(\mathrm{vec}^{\rm T}\left\lbrace\mathrm{vec}\left\lbrace\frac{\partial\overline{\gamma}_{\mathcal{R}_k}}{\partial\boldsymbol{\mathcal{F}}}\right\rbrace\right\rbrace\right) \end{array}\right]\nonumber\\
=&\left[ \begin{array}{ccc}
\mathbf{0}_{N^2\times N^2} &
\mathbf{0}_{N^2\times N^2}\\
\frac{\partial}{\partial\boldsymbol{\mathcal{F}}}\left(\mathrm{vec}^{\rm T}\left\lbrace\mathrm{vec}\left\lbrace\frac{\partial\overline{\gamma}_{\mathcal{R}_k}}{\partial\boldsymbol{\mathcal{F}}}\right\rbrace\right\rbrace\right) & \mathbf{0}_{N^2\times N^2} \end{array}\right],
\end{align} 
using the two underlying first order partial derivatives:  
\begin{subequations}
	\begin{align}	
	\frac{\partial\overline{\gamma}_{\mathcal{R}_k}}{\partial\boldsymbol{\mathcal{F}}}=&\,\Bigg[-\frac{\overline{\gamma}_{\mathcal{R}_k}\sum_{i\in\mathcal{M}_k}\alpha_i\left|\mathbf{g}_{k}^{\rm H}\,\mathbf{h}_{i}\right|^2\,{\mathbf{h}_{i}^*\,\mathbf{h}_{i}^{\rm T}}}{\sum_{i\in\mathcal{M}_k}\alpha_i\left|\mathbf{g}_{k}^{\rm H}\,\mathbf{h}_{i}\right|^2\,
		{\mathbf{h}_{i}^{\rm T}\boldsymbol{\mathcal{F}}\,\mathbf{h}_{i}^*}+\sigma_{\mathrm{w}_{\mathcal{R}}}^2\norm{\mathbf{g}_{k}}^2}\nonumber\\
	 &\qquad+\frac{\overline{\gamma}_{\mathcal{R}_k}{\mathbf{h}_{k}^*\,\mathbf{h}_{k}^{\rm T}}}{{\mathbf{h}_{k}^{\rm T}\boldsymbol{\mathcal{F}}\,\mathbf{h}_{k}^*}}\Bigg]\in\mathbb{C}^{N\times N} ,\,\forall k\in\mathcal{M},
	\end{align}
	\begin{equation}
	\frac{\partial\overline{\gamma}_{\mathcal{R}_k}}{\partial\boldsymbol{\mathcal{F}}^*}=\mathbf{0}_{N\times N},\quad\forall k\in\mathcal{M}.
	\end{equation}
\end{subequations}
As all the principal minors of $\mathbb{H}\big(\overline{\gamma}_{\mathcal{R}_k}\big)$, of any order $k\le2N^2$ are zero, we can observe that it is  negative-semidefinite, which implies that the backscattered SINR $\overline{\gamma}_{\mathcal{R}_k}$ for each  $\mathcal{T}_k$ is concave in $\boldsymbol{\mathcal{F}}$. This concavity of $\overline{\gamma}_{\mathcal{R}_k}$ in $\boldsymbol{\mathcal{F}}$ can also be more simply realized in scalar form, where each SINR term $\left(\text{say, } \mathcal{G}\left(x\right)\triangleq\frac{a_1\,x}{a_2\,x+a_3}\right)$, involving \textit{the ratio of  univariate positive linear and  affine functions}, is a strictly-concave function $\left(\text{because  }\frac{\partial^2\mathcal{G}}{\partial x^2}=-\frac{2 a_1 a_2 a_3}{\left(a_2\,x+a_3\right)^3}\le0,\forall a_1,a_2,a_3\ge0\right)$.  Lastly, noting that the concavity is preserved under a concave monotonically increasing transformation~\cite[eq. (3.10)]{boyd}, like  $\overline{\mathrm{R}}_k\triangleq\log_2\left(1+\overline{\gamma}_{\mathcal{R}_k}\right)$ as a function of $\overline{\gamma}_{\mathcal{R}_k}$, we note that each backscattered-throughput $\overline{\mathrm{R}}_k$, along with their sum $\overline{\mathrm{R}}_{\mathrm{S}}=\sum_{k\in\mathcal{M}}\,\log_2\left(1+\overline{\gamma}_{\mathcal{R}_k}\right)$, are all concave in  $\boldsymbol{\mathcal{F}}$.

\setcounter{equation}{0}
\setcounter{figure}{0}
\renewcommand{\theequation}{C.\arabic{equation}}
\renewcommand{\thefigure}{C.\arabic{figure}}
\section{Proof of Lemma~\ref{lem:high-opt}:   Concavity of $\overline{\mathrm{R}}_{\mathrm{S_H}}$ in  $\boldsymbol{\mathcal{F}}$}\label{app:high-opt}  
Firstly, we note that regardless of the value of $\boldsymbol{\mathcal{F}}$, $\overline{\mathrm{R}}_{\mathrm{S_H}}$ is monotonically increasing in each $\alpha_k,\,\forall k\in\mathcal{M}.$ So, optimal BC under high-SNR scenario is given by $\alpha_{\mathrm{H}_k}=\alpha_{\max},\forall k\in\mathcal{M}.$ Now, with both $\boldsymbol{\alpha}=\boldsymbol{\alpha}_{\rm H}\triangleq\alpha_{\max}\mathbf{1}_{M\times1}$ and $\mathbf{G}=\mathbf{G}_{\rm H}$ obtained, we next focus on obtaining the optimal precoding vector $\mathbf{f}$ in high-SNR regime. We start with showing that $\mathcal{O}_{\mathrm{H}}$  involves the maximization of $\overline{\mathrm{R}}_{\mathrm{S_H}}$, i.e., the sum of $M$ concave functions $\overline{\mathrm{R}}_{\mathrm{H}_k}\triangleq\log_2\left(1+\alpha_k\,\widetilde{\gamma}_{\mathrm{g}_k}{\mathbf{h}_{k}^{\rm T}\boldsymbol{\mathcal{F}}\,\mathbf{h}_{k}^*} \right)$ over the variable $\boldsymbol{\mathcal{F}}$. Here, the concavity of each throughput term $\overline{\mathrm{R}}_{\mathrm{H}_k}$ is proved by investigating the combined Hessian matrix $\mathbb{H}\big(\overline{\mathrm{R}}_{\mathrm{H}_k}\big)$ of $\overline{\mathrm{R}}_{\mathrm{H}_k}$ with respect to $\boldsymbol{\mathcal{F}}$ as defined below:
\begin{align}
\mathbb{H}\big(\overline{\mathrm{R}}_{\mathrm{H}_k}\big)=&\,\left[ \begin{array}{ccc}
\boldsymbol{\mathcal{H}}_{\boldsymbol{\mathcal{F}}\boldsymbol{\mathcal{F}}^*}\big(\overline{\mathrm{R}}_{\mathrm{H}_k}\big)
&\boldsymbol{\mathcal{H}}_{\boldsymbol{\mathcal{F}}^*\boldsymbol{\mathcal{F}}^*}\big(\overline{\mathrm{R}}_{\mathrm{H}_k}\big) \\
\boldsymbol{\mathcal{H}}_{\boldsymbol{\mathcal{F}}\boldsymbol{\mathcal{F}}}\big(\overline{\mathrm{R}}_{\mathrm{H}_k}\big)
&\boldsymbol{\mathcal{H}}_{\boldsymbol{\mathcal{F}}^*\boldsymbol{\mathcal{F}}}\big(\overline{\mathrm{R}}_{\mathrm{H}_k}\big)  \end{array} \right]\nonumber\\
=&\,\left[ \begin{array}{ccc}
\mathbf{0}_{N^2\times N^2} &
\mathbf{0}_{N^2\times N^2}\\
\frac{\partial}{\partial\boldsymbol{\mathcal{F}}}\left(\mathrm{vec}^{\rm T}\left\lbrace\mathrm{vec}\left\lbrace\frac{\partial\overline{\mathrm{R}}_{\mathrm{H}_k}}{\partial\boldsymbol{\mathcal{F}}}\right\rbrace\right\rbrace\right) & \mathbf{0}_{N^2\times N^2} \end{array} \right],
\end{align} 
where $\frac{\partial\overline{\mathrm{R}}_{\mathrm{H}_k}}{\partial\boldsymbol{\mathcal{F}}}=\frac{\alpha_k\,\widetilde{\gamma}_{\mathrm{g}_k}{\mathbf{h}_{k}^*\,\mathbf{h}_{k}^{\rm T}} }{\left(1+\alpha_k\,\widetilde{\gamma}_{\mathrm{g}_k}{\mathbf{h}_{k}^{\rm T}\boldsymbol{\mathcal{F}}\,\mathbf{h}_{k}^*}\right)\ln\left(2\right)}$ and  $\frac{\partial\overline{\mathrm{R}}_{\mathrm{H}_k}}{\partial\boldsymbol{\mathcal{F}}^*}=\mathbf{0}_{N\times N}$. 
As all the principal minors of $\mathbb{H}\big(\overline{\mathrm{R}}_{\mathrm{H}_k}\big)$, of any order $k\le2N^2$, are zero, we can observe that it is negative-semidefinite. Hence, $\overline{\mathrm{R}}_{\mathrm{H}_k}$ for   $\mathcal{T}_k$ is concave in $\boldsymbol{\mathcal{F}}$. This result can also be  interpreted from the fact that $\overline{\mathrm{R}}_{\mathrm{H}_k}$ is a \textit{concave monotonically increasing}  transformation of an affine function of $\boldsymbol{\mathcal{F}}$.



\end{document}